\RequirePackage{tikz}

\documentclass[pdflatex,sn-basic]{sn-jnl}

\jyear{2021}%

\theoremstyle{thmstyleone}%
\theoremstyle{thmstyletwo}%

\theoremstyle{thmstylethree}%

\usepackage{subcaption}
\usepackage{tikzscale}

\usepackage{pgfplots}
\pgfplotsset{compat=1.18}

\newcommand{\euler}{e}

\newcommand{\imag}{\mathrm{i}}

\renewcommand{\Vec}[1]{\boldsymbol{#1}}

\newcommand{\Tensor}[1]{\mathsf{#1}}

\newcommand{\FT}[2]{\mathcal{F}\left\lbrace{#1}\right\rbrace\!\left({#2}\right)}
\newcommand{\IFT}[2]{\mathcal{F}^{-1}\left\lbrace{#1}\right\rbrace\!\left({#2}\right)}

\newcommand{\superficialavg}[1]{\left\langle#1\right\rangle_{\mathrm{s}}}
\newcommand{\intrinsicavg}[1]{\left\langle#1\right\rangle_{\mathrm{i}}}

\newcommand{\porosity}{\epsilon}

\newcommand{\permeability}{K}

\newcommand{\statictortuosity}{\alpha_{0}}

\newcommand{\tortuosity}{\alpha_{\infty}}

\newcommand{\Rey}{\mbox{\textit{Re}}} 
\newcommand{\Wo}{\mbox{\textit{Wo}}}  
\newcommand{\Hg}{\mbox{\textit{Hg}}}  

\begin{document}

\title[Assessment of models for nonlinear oscillatory flow \dots]{Assessment of models for nonlinear oscillatory flow through a hexagonal sphere pack}


\author*[1,2]{\fnm{Lukas} \sur{Unglehrt}}\email{lukas.unglehrt@tum.de}

\author[1]{\fnm{Michael} \sur{Manhart}}\email{michael.manhart@tum.de}

\affil[1]{\orgdiv{Professorship for Hydromechanics}, \orgname{Technical University of Munich}, \orgaddress{\street{Arcisstr. 21}, \postcode{80333} \city{Munich}, \country{Germany}}}

\affil[2]{\orgname{Unglehrt GmbH \& Co. KG}, \orgaddress{\street{Allg\"auer Str. 31}, \postcode{87700} \city{Memmingen}, \country{Germany}}}


\abstract{

We review models for unsteady porous media flow in the volume-averaging framework and we discuss the theoretical relations between the models and the definition of the model coefficients (and the uncertainty therein). The different models are compared against direct numerical simulations of oscillatory flow through a hexagonal sphere pack. The model constants are determined based on their definition in terms of the Stokes flow, the potential flow and steady nonlinear flow. Thus, the discrepancies between the model predictions and the simulation data can be attributed to shortcomings of the models' parametrisation. \\

We found that an extension of the dynamic permeability model of Pride et al. [Physical Review B 47(9), 1993] with a Forchheimer-type nonlinearity performs very well for linear flow and for nonlinear flow at low and medium frequencies, but the Forchheimer term with a coefficient obtained from the steady-state overpredicts the nonlinear drag at high frequencies. The model reduces to the unsteady Forchheimer equation with an acceleration coefficient based on the static viscous tortuosity for low frequencies. \\

The unsteady Forchheimer equation with an acceleration coefficient based on the high frequency limit of the dynamic tortuosity has large errors for linear flow at medium and high frequencies, but low errors for nonlinear flow at all frequencies. This is explained by an error cancellation between the inertial and the nonlinear drag.
}




\maketitle

\section*{Article highlights}

\begin{itemize}
    \item We review models for unsteady porous media flow in the volume-averaging framework  and discuss their relationships.
    \item The model predictions are compared to direct numerical simulations of oscillatory flow through a hexagonal sphere pack.
    \item Accurate models exist for the linear drag, but the Forchheimer term overpredicts the nonlinear drag at high frequencies.
\end{itemize}

\section{Introduction}

Unsteady flow through porous media occurs in a variety of environmental, engineering and industrial applications. For instance, wave-induced flow through coral reefs \citep{Lowe.2008},  breakwaters \citep{vanGent.1994,Losada.1995,Hall.1995,Muttray.2000} and marine sediment \citep{Gu.1991} has been described using the theory of porous media. Oscillatory flow is also present in combustion engines where porous media burners could lead to emission reductions \citep{Aboujafari.2022}. Moreover, porous media have been used as regenerator-type heat exchangers in Stirling engines \citep{Simon.1988,Trevizoli.2016}. In chemical reactors, pulsating flow could be used to enhance mixing and mass transfer \citep{Ni.2003} or to separate substances \citep{Graham.2002}. The increasing use of wind energy may also lead to an interest in processes with an intermittent energy supply. Furthermore, the understanding of transient flow behaviour is important for the safety design of nuclear pebble bed reactors \citep{Andreades.2014}. Finally, \citet{Kahler.2019} suggested to use pulsating flow to accelerate groundwater remediation.

The pore scale flow through a porous medium is governed by the laws of continuum mechanics, i.e. the Navier-Stokes equations, or statistical mechanics, i.e. the Boltzmann equation. However, a direct solution of the pore scale flow is often computationally demanding and knowledge of the pore geometry is often not available. Therefore, coarse-grained descriptions have been developed, made possible by the scale separation between the pore size and the extent of the porous medium. These coarse-grained descriptions are based, for example, on the volume-averaging framework \citep{Whitaker.1967} or homogenisation theory \citep{Ene.1975}. A comparison between these approaches can be found in \citep{Davit.2013}.  Descriptions based on the volume-averaging approach have been used for example by \citet{Gu.1991}, who studied gravity waves over a porous seabed, by \citet{vanGent.1994}, who investigated wave transmission through dikes and breakwaters, by \citet{Breugem.2006}, who studied turbulent channel flow over porous media, and by \citet{Iliuta.2016,Iliuta.2017}, who simulated oscillating packed-bed reactors for offshore applications. We now give a brief outline of the volume-averaging method. The macroscopic quantities are obtained by performing a (weighted) local average of the quantities defined at the pore scale over a representative volume element. For example, the superficial velocity and the macroscopic pressure are defined as
\begin{subequations}
    \begin{align}
        \superficialavg{\Vec{u}} &= \frac{1}{V}\int_{V_{\mathrm{f}}} \Vec{u}\,\mathrm{d}V \\
        \intrinsicavg{p} &= \frac{1}{V_{\mathrm{f}}}\int_{V_{\mathrm{f}}} p\,\mathrm{d}V \,,
    \end{align}
\end{subequations}
where $\superficialavg{.}$ and $\intrinsicavg{.}$ denote the superficial and intrinsic volume average, respectively, and $V$ and $V_{\mathrm{f}}$ are the volumes of the representative volume element and of the fluid contained therein. These quantities are governed by the volume-averaged Navier-Stokes (VANS) equations \citep{Whitaker.1996}
\begin{subequations}
    \begin{align}
        \Vec{\nabla}\cdot\superficialavg{\Vec{u}} &= 0 \label{eq:avg_continuity} \\
        \rho\,\frac{\partial\!\superficialavg{\Vec{u}}}{\partial t}&=\underbrace{-\frac{1}{V}\int_{A_{\mathrm{fs}}} \tilde{p}\,\Vec{n}\,\mathrm{d}A}_{\text{pressure drag}} \underbrace{-\frac{1}{V} \int_{A_{\mathrm{fs}}} \Vec{\tau}_\mathrm{w}\,\mathrm{d}A}_{\text{friction drag}} - \porosity\,\Vec{\nabla}\!\intrinsicavg{p} \label{eq:avg_momentum_equation}
    \end{align}
\end{subequations}
where $\tilde{p}=p-\intrinsicavg{p}$, $\porosity=V_{\mathrm{f}}/V$ is the porosity and $A_{\mathrm{fs}}$ is the fluid-solid interface. Other works, e.g. \citep{Hsu.1990}, have also included convective and diffusive terms in the superficial velocity, but these can be generally neglected if the scale separation between the pore scale and the averaging scale is large enough \citep{Whitaker.1986,Whitaker.1996}. In the VANS equations the pressure drag and the friction drag are unclosed with respect to the superficial velocity and the macroscopic pressure gradient. In the approach of \citet{Whitaker.1986,Whitaker.1996}, the pore scale velocity and pressure are expressed in terms of the superficial velocity by linear mappings. These mappings take the form of tensor and vector fields that satisfy a boundary value problem in the pore space of a representative volume element (``closure problem''). When the mappings are substituted into the pressure drag and friction drag terms, the drag terms take the form of a product of the inverse of a permeability-like tensor with the superficial velocity. In steady linear flow, the closure problem depends only on the geometry of the pore space \citep{Whitaker.1986}, and consequently the permeability tensor of the porous medium is independent of the flow history and the fluid properties. In nonlinear flow, however, the closure problem depends on the velocity field on the pore scale \citep{Whitaker.1996} and the permeability-like tensor depends on the flow history and the fluid properties. Thus, a direct numerical simulation of the pore scale flow in a representative volume element is required to solve the closure problem. For three-dimensional porous media, these simulations require a large computational effort. Solving the pore scale problem can be avoided by parametrising the nonlinear drag directly in terms of the superficial velocity and the macroscopic pressure gradient. This is based on experiments or numerical simulations of representative volume elements that are performed before solving the macroscale problem. In the following, we refer to these parametrisations as \emph{models}.

The aim of the present work is to compare different models for unsteady porous media flow for the special case of oscillatory flow. We address the following research questions: What is the domain of validity for the different models? How should the model coefficients be chosen? How can the models be improved?

In this contribution, we first describe some of the prominent models that are available in the literature and discuss their interrelations. We then compare the predictions of the different models with a high fidelity direct numerical simulation dataset of oscillatory flow through a hexagonal sphere pack. We make the assumption of constant model coefficients that are defined in terms of the linear and the steady state behaviour and represent properties of the porous medium geometry. For the present flow configuration this information is available with a high fidelity based on the works of \citet{Zhu.2016a,Sakai.2020,Unglehrt.2022}. This allows us to test the actual predictive capabilities of the models with a negligible ambiguity in the values of the model coefficients. The errors of the predictions therefore represent a shortcoming of the model and suggest the need for a different parametrisation.

\section{Review of models for porous media flow}

In this section, we give an overview of some of the common models for porous media flow. As discussed in the introduction, we use the term \emph{model} to refer to a parametrisation of the drag in the volume-averaged momentum equation.

\subsection{Models for linear flow}
\label{sec:linear_models}

\subsubsection{Darcy equation}

In steady conditions, linear flow can be described using the Darcy equation \citep{Darcy.1856}
\begin{equation}
	-\Vec{\nabla}\!\intrinsicavg{p}=\frac{\mu}{\permeability} \superficialavg{\Vec{u}}
	\label{eq:Darcy_equation}
\end{equation}
The Darcy equation relates the pressure gradient and the superficial velocity linearly using the dynamic viscosity $\mu$ and the permeability $\permeability$. The permeability has units of length squared and is a pure function of the pore geometry. For a given pore geometry, it can be computed directly from the solution to the Stokes equations for a given pore geometry or from empirical correlations, e.g. the Kozeny-Carman equation.

\subsubsection{Unsteady Darcy equation}

The unsteady Darcy equation arises from the volume-averaged Navier-Stokes equations \eqref{eq:avg_momentum_equation} if the quasi-steady closure with Darcy's law is employed for the drag forces
\begin{equation}
    \rho\,\frac{\mathrm{d}\!\superficialavg{\Vec{u}}}{\mathrm{d} t}= - \porosity\,\Vec{\nabla}\!\intrinsicavg{p} - \frac{\porosity\,\mu}{\permeability}\superficialavg{\Vec{u}} \,.
\end{equation}
The solutions to this equation relax to Darcy's law \eqref{eq:Darcy_equation} with a time constant $\tau_{\mathrm{vans}}=\permeability/(\porosity\,\nu)$. This model was applied by \citet{Kuznetsov.2006} to pulsating and by \citet{Wang.2008} to transient porous media flow, respectively; in these works the unsteady Darcy equation was extended by the Brinkman term in order to account for wall boundary conditions.

As discussed for example by \citet{Nield.1991}, an a priori unknown acceleration coefficient needs to be introduced in front of the time derivative. Based on a virtual mass analogy, \citet{Sollitt.1972} and \citet{Gu.1991} used a factor $1+C_{\mathrm{M}}\frac{1-\porosity}{\porosity}$ in front of the acceleration term which results in a time constant $\tau_{\mathrm{vm}} = \left[1+C_{\mathrm{M}}\frac{1-\porosity}{\porosity}\right] \permeability/(\porosity\,\nu)$. Finally, \citet[eq. (20) and (21)]{Hill.2001} and \citet{Zhu.2014} derived another time constant $\tau_{\mathrm{en}}=\statictortuosity\,\tau_{\mathrm{vans}}$ from the volume-averaged kinetic energy equation assuming self-similar velocity profiles and a quasi-steady dissipation rate where $\statictortuosity = \intrinsicavg{\Vec{u}_{\text{Stokes}}^2}/\intrinsicavg{\Vec{u}_{\text{Stokes}}}^2$ is the static viscous tortuosity that is defined in terms of the velocity field of the Stokes flow \citep[p.156f]{Lafarge.1993}. In the literature, the static viscous tortuosity  $\statictortuosity$  has been referred to as ``inertial factor'' \citep{Norris.1986}, ``acceleration coefficient'' \citep{Nield.1991}, ``low-frequency limit of the dynamic tortuosity'' \citep{Champoux.1991}, ``low frequency viscous [...] [equivalent] of the tortuosity [$\tortuosity$]'' \citep{Cortis.2002}, ``viscous tortuosity'' \citep{Kergomard.2013}, ``time scale ratio'' \citep{Zhu.2016a}, ``static viscous tortuosity'' \citep{Roncen.2018}.

It was shown by \citet{Zhu.2014,Zhu.2016a} that the unsteady Darcy equation with the time constant $\tau_{\mathrm{en}}$, i.e.
\begin{equation}
	\rho \,\statictortuosity\, \frac{\mathrm{d}\!\superficialavg{\Vec{u}}}{\mathrm{d}t} = -\porosity\,\Vec{\nabla}\!\intrinsicavg{p} - \frac{\porosity\,\mu}{\permeability}\superficialavg{\Vec{u}}
	\label{eq:unsteady_Darcy_equation} \,,
\end{equation}
is the appropriate choice for transient flow and low frequency oscillatory flow whereas for high frequency oscillatory flow a different time constant needs to be employed. This will be further discussed in section \ref{sec:discussion_of_models}.

\subsubsection{Dynamic permeability models}
\label{sec:dynamic_permeability_models}

Linear oscillatory flow through porous media has been studied extensively in acoustics. \citet{Johnson.1987} proposed an important family of models, the dynamic permeability. These are based on a generalisation of the Darcy equation \eqref{eq:Darcy_equation} in the frequency domain \footnote{
	We define the Fourier transform of a function $g(t)$ and the inverse Fourier transform of a function $\hat{g}(\omega)$  as
	\begin{align}
		\FT{g(t)}{\omega} &= \int_{-\infty}^{\infty} g(t) \euler^{-\imag \omega t}\,\mathrm{d}t \\
		\IFT{\hat{g}(\omega)}{t} &= \frac{1}{2\pi}\int_{-\infty}^{\infty} \hat{g}(\omega) \euler^{\imag \omega t}\,\mathrm{d}\omega
	\end{align}
}
\begin{equation}
    \FT{\superficialavg{\Vec{u}}}{\omega}= -\frac{\hat{\permeability}(-\omega)}{\mu} \FT{\Vec{\nabla}\!\intrinsicavg{p}}{\omega}
    \label{eq:Darcy_equation_Fourier_space}
\end{equation}
where the function $\hat{\permeability}(\Omega)$ is referred to as dynamic permeability and corresponds to the frequency response function in linear systems theory. 

\citet{Johnson.1987} derived high-frequency asymptotics from boundary layer theory (cf. \citet{Cortis.2003}) and proposed a model for the entire frequency range by blending the high-frequency asymptotics with Darcy's law at low frequencies. The model by \citet{Johnson.1987} is given in the notation of \citet{Pride.1993} for a complex frequency $\Omega$ as
\begin{equation}
    \hat{\permeability}(\Omega) = \frac{\permeability}{\sqrt{1-\imag P \frac{\Omega}{\Omega_0}}-\imag\frac{\Omega}{\Omega_0}}\,,
    \label{eq:Johnson_model}
\end{equation}
with the frequency $\Omega_0 = \porosity\,\nu/(\tortuosity\,\permeability)$ and the dimensionless parameter $P = 4\,\tortuosity\,\permeability/(\porosity\,\Lambda^2)$. The transition frequency $\Omega_0$ ``separates viscous-force-dominated flow from inertial-force flow'' \citep{Pride.1993}. The original parameters introduced by \citet{Johnson.1987} are the high-frequency limit of the dynamic tortuosity $\tortuosity$ (identical to the ratio $\intrinsicavg{\Vec{u}^2}/\intrinsicavg{\Vec{u}}^2$ in potential flow) and a length-scale $\Lambda$. Both parameters can be computed in terms of the potential flow solution. \citet{Smeulders.1992} presented a derivation of the model of \citet{Johnson.1987} from first principles using homogenisation theory. \citet{Chapman.1992} compared the model predictions to numerical solutions of the unsteady Stokes equations obtained with a least-squares collocation approach and found good agreement. Similarly, in \citep{Unglehrt.2022} we observed excellent agreement of the model with our simulations. For large frequencies, the model  \eqref{eq:Johnson_model} reduces to
\begin{equation}
    \hat{\permeability}(\Omega) = \frac{\permeability}{\sqrt{-\imag P \frac{\Omega}{\Omega_0}}-\imag\frac{\Omega}{\Omega_0}}
    \label{eq:Johnson.1987_high_frequency_aymptotics}
\end{equation}
Using the inverse Fourier transform of equations \eqref{eq:Darcy_equation_Fourier_space} and \eqref{eq:Johnson.1987_high_frequency_aymptotics}, we can write the momentum equation in the time domain \citep{Turo.2013}
\begin{equation}
    \rho\,\frac{\mathrm{d}\!\superficialavg{\Vec{u}}}{\mathrm{d}t} = -\frac{2\rho \sqrt{\nu}}{\Lambda} \int_{-\infty}^{t} \frac{\mathrm{d}\!\superficialavg{\Vec{u}}}{\mathrm{d}\tau}\frac{1}{\sqrt{\pi (t-\tau)}}\,\mathrm{d}\tau- \frac{\porosity}{\tortuosity}\Vec{\nabla}\!\intrinsicavg{p}\,,
\end{equation}
where the integral term corresponds to the Caputo fractional derivative of order $\frac{1}{2}$. This term originates from boundary layer theory and is analogous to the Basset history term in the solution for flow around a sphere \citep{Turo.2013}. Consequently, the flow state in the dynamic permeability model \eqref{eq:Johnson_model} requires the specification of the entire history of $\superficialavg{\Vec{u}}\!(t)$. This is in contrast to the unsteady Darcy equation where the flow state is completely specified by the instantaneous value of $\superficialavg{\Vec{u}}$.

In order to improve the low-frequency behaviour, \citet{Pride.1993} devised an extended model
\begin{equation}
	\hat{\permeability}(\Omega) = \frac{\permeability}{\left(1-\frac{P}{2\beta}+\sqrt{\frac{P^2}{4\beta^2}-\imag P \frac{\Omega}{\Omega_0}}\right)-\imag\frac{\Omega}{\Omega_0}}\,.
\label{eq:Pride_model}
\end{equation}
The model of \citet{Johnson.1987} is obtained in the special case $P=2\beta$ where the additional non-dimensional parameter $\beta = \frac{\statictortuosity}{\tortuosity}-1$ is defined in terms of the static viscous tortuosity $\statictortuosity$ and the high-frequency limit of the dynamic tortuosity $\tortuosity$. It is therefore a measure for the difference in time scales of Stokes flow and potential flow.

Another extension of the model of \citet{Johnson.1987} was developed by \citet{Champoux.1991} to represent thermal dissipation effects occuring for an ideal gas. As we restrict our analyses to incompressible flow, we will not discuss this any further. A comprehensive discussion of the dynamic permeability models (also known as \textit{equivalent fluid model}) can be found in \citet{Lafarge.2009}.

The dynamic permeability models can also be written in the time domain by performing an inverse Fourier transform. For example, a time domain formulation of the model of \citet{Johnson.1987} (equation \ref{eq:Johnson_model}) was given by \citet{Umnova.2009}. In the same way, we obtained the time domain formulation of the more general model of \citet{Pride.1993} (equation \ref{eq:Pride_model}):
\begin{equation}
	\begin{split}
		\rho\,\frac{\mathrm{d}\!\superficialavg{\Vec{u}}}{\mathrm{d}t} =& -\frac{\porosity}{\tortuosity} \Vec{\nabla}\!\intrinsicavg{p} - \left(\frac{\porosity}{\tortuosity}\frac{\mu}{\permeability}-\frac{2\mu}{\Lambda^2\,\beta}\right)\superficialavg{\Vec{u}} 
		\\
		&-\frac{2 \rho \sqrt{\nu}}{\Lambda}\int_{-\infty}^{t} \left(\frac{\nu \superficialavg{\Vec{u}}}{\Lambda^2\,\beta^2}+\frac{\mathrm{d}\!\superficialavg{\Vec{u}}}{\mathrm{d}\tau}\right)\frac{\euler^{- \frac{\nu (t-\tau)}{\Lambda^2\,\beta^2}  }}{\sqrt{\pi(t-\tau)}}\,\mathrm{d}\tau \,.
	\end{split}
	\label{eq:Pride_model_time_domain} 
\end{equation}
Note that we have inserted the definitions of the parameters $P$ and $\Omega_0$ in order to ease the comparison with the other time domain models. The kernel in the convolution integral represents an exponential damping of the fractional derivative kernel and reduces the weight on the history further in the past.

\subsection{Models for nonlinear flow}

\subsubsection{Darcy equation with cubic correction}
For steady flow at small $\Rey$, \citet{Mei.1991} showed using a homogenization theory approach that Darcy's law needs to be corrected with a cubic term
\begin{equation}
-\Vec{\nabla}\!\intrinsicavg{p}= \frac{\mu}{\permeability}\superficialavg{\Vec{u}} + \frac{\mu\,b}{\permeability}\superficialavg{\Vec{u}}^2\superficialavg{\Vec{u}}
\end{equation}
where $b$ is a non-negative coefficient with the units $T^2/L^2$. \citet{Koch.1997} simulated flow through periodic and random arrays of cylinders and confirmed that the first correction to Darcy's law is cubic with respect to the bulk velocity. \citet{Firdaouss.1997} showed that the cubic correction also holds for anisotropic media provided that the modulus of the bulk velocity does not change under a reversal of the driving force.  \citet{Hill.2001} investigated flow through random and ordered arrays of spheres using DNS and concluded that ``[a]t all solid volume fractions, the first inertial contribution to the non-dimensional drag force was found to be proportional to the square of the Reynolds number, as predicted by the theory of Mei \& Auriault.''. Numerical investigations of \citet{Lasseux.2011} suggest that the cubic correction is valid for permeability Reynolds numbers $\Rey_{\permeability} =\left\vert\superficialavg{\Vec{u}}\right\vert \sqrt{\permeability}/\nu$ up to about $0.1-0.8$ depending on the porosity and the pore geometry. 

\subsubsection{Forchheimer equation}
For nonlinear flow at higher Reynolds numbers, \citet{Forchheimer.1901} proposed the empirical equation
\begin{equation}
	-\Vec{\nabla}\!\intrinsicavg{p}=a \superficialavg{\Vec{u}} + b  \left\vert\superficialavg{\Vec{u}}\right\vert \superficialavg{\Vec{u}}
	\label{eq:Forchheimer_equation}
\end{equation}
where the coefficients $a$ and $b$ have units of $M/(L^3\,T)$ and $M/L^4$, respectively. For packed beds of spheres, comprehensive empirical correlations for these coefficients were first given by \citet{Ergun.1952}. Updated forms of the correlations have been given e.g. by \citet{Macdonald.1979}. When the flow in the pore space becomes turbulent, a different set of coefficients should be chosen  \citep{Burcharth.1995}; this results in a piecewise description of the drag.

\subsubsection{Unsteady Forchheimer equation}
\label{sec:unsteady_Forchheimer_equation}

For the description of unsteady porous media flow, \citet{Polubarinova-Kochina.1962} proposed to extend the Forchheimer equation \eqref{eq:Forchheimer_equation} with an acceleration term
\begin{equation}
    -\Vec{\nabla}\!\intrinsicavg{p}= a\superficialavg{\Vec{u}} + b\left\vert\superficialavg{\Vec{u}}\right\vert \superficialavg{\Vec{u}}+c \frac{\mathrm{d}\!\superficialavg{\Vec{u}}}{\mathrm{d}t} \,.
    \label{eq:unsteady_Darcy-Forchheimer_equation}
\end{equation}
\citet{Sollitt.1972} derived a parametrisation of this equation where $a$ and $b$ were chosen according to the steady state equation by \citet{Ward.1964} and the form of the acceleration coefficient $c$ was determined based on a virtual mass argument. Their equation reads
\begin{subequations}
    \begin{equation}
    \rho\,S\, \frac{\mathrm{d}\!\superficialavg{\Vec{u}}}{\mathrm{d}t} =-\porosity\,\Vec{\nabla}\!\intrinsicavg{p}-\frac{\porosity\,\mu}{\permeability}\superficialavg{\Vec{u}}-\rho\,\frac{\porosity\,C_{\mathrm{f}}}{\sqrt{\permeability}} \left\vert\superficialavg{\Vec{u}}\right\vert \superficialavg{\Vec{u}}
\end{equation}
where the ``inertial coefficient'' is defined as
\begin{equation}
    S = 1+\frac{1-\porosity}{\porosity}C_{\mathrm{M}} \,.
\end{equation}
\end{subequations}
However, \citet{Sollitt.1972} considered the coefficient $C_{\mathrm{M}}$, which represents the virtual mass of the solid grains, as unknown and set it to zero. In contrast, the experimental investigation of \citet{Gu.1991} resulted in $C_{\mathrm{M}}= 0.46$ for gravel beds with a porosity between $0.35$ and $0.38$. Also, other parametrisations of the acceleration coefficient have been given in later works \citep{Burcharth.1995}.

Another choice for the coefficient $c$ was proposed by \citet{Zhu.2016} who suggested to use the time constant $\tau_{\mathrm{en}}=\statictortuosity\,\permeability/(\porosity \nu)$ in direct analogy to the unsteady Darcy equation \eqref{eq:unsteady_Darcy_equation}.

\citet{Burcharth.1995} state that ``the coefficients [$b$ and $c$] are not constants and should in principle be treated as instantaneous values, even for oscillatory flow conditions''. For example, \citet{Hall.1995} calculated the instantaneous values of the coefficients $a$, $b$ and $c$ from oscillatory flow data. However, no explicit parametrisation of the constants was determined. On the other hand, \citet{vanGent.1993} considered the coefficients as constants for each flow case and found frequency-dependent correlations for the coefficients $b$ and $c$. However, this kind of parametrisation is specific to the flow case and is not generally applicable \citep{Burcharth.1995}.

Furthermore, it is unclear how the change of the drag behaviour due to the transition to turbulence (see \citet{Burcharth.1995}) could be incorporated into the unsteady Forchheimer equation. In oscillatory flow the critical Reynolds number of transition depends on the frequency (see section \ref{sec:description_of_the_simulation_dataset} or the estimations of \citet{Gu.1991}), ruling out the straightforward way of changing the coefficients $a$, $b$, and $c$ depending on the instantaneous Reynolds number.

\subsubsection{Extended dynamic permeability model}

\citet{Turo.2013} proposed a model for unsteady nonlinear porous media flow. The model combines the time domain formulation of a dynamic permability model with a Forchheimer-type quadratic term:
\begin{equation}
	\begin{split}
        \rho \frac{\mathrm{d}\!\superficialavg{\Vec{u}}}{\mathrm{d}t}= &- \frac{\porosity}{\tortuosity} \Vec{\nabla}\!\intrinsicavg{p} - \frac{\porosity}{\tortuosity}\frac{\mu}{\permeability} \left(1+\xi \vert\!\superficialavg{\Vec{u}}\vert\right) \superficialavg{\Vec{u}} \\
        &- \frac{2 \rho \sqrt{\nu}}{\Lambda}  \int_{-\infty}^{t} \frac{\mathrm{d}\!\superficialavg{\Vec{u}}}{\mathrm{d}\tau} \frac{1}{\sqrt{\pi (t-\tau)}} \,\mathrm{d}\tau \,.
	\end{split}
	\label{eq:Turo_model}
\end{equation}
The parameter $\xi \,\,[T/L]$ describes the nonlinearity and is related to the Forchheimer coefficient $b$ as $\xi=b \permeability/\mu$. This model can be interpreted as an additive combination of the drag due to the boundary layers (corresponding to the high-frequency asymptotics \eqref{eq:Johnson.1987_high_frequency_aymptotics} of \citet{Johnson.1987}) and of the drag in the steady state described by the Forchheimer equation \eqref{eq:Forchheimer_equation}.

\subsection{Discussion}
\label{sec:discussion_of_models}

\subsubsection{Relations among the linear models}
\label{sec:relations_among_the_linear_models}

\begin{figure}
    \centering
    \includegraphics[width=\textwidth]{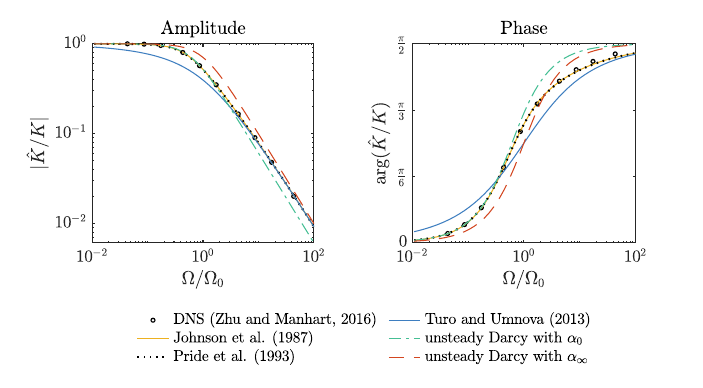}
    
    \caption{Comparison of different linear models with the direct numerical simulations of \citet{Zhu.2016a}. Amplitude (left) and phase (right) of the dynamic permeability $\hat{\permeability}$ normalised with the permeability. The dynamic permeability functions of the models are given by the equations \eqref{eq:Johnson_model}, \eqref{eq:Pride_model}, \eqref{eq:Turo_Umnova_frequency_domain} and \eqref{eq:unsteady_Darcy_frequency_domain}, respectively.}
    \label{fig:dyn_permeability}
\end{figure}

The linear models presented in section \ref{sec:linear_models} can be understood as various special cases of the dynamic permeability model of \citet{Pride.1993} given in equation \eqref{eq:Pride_model}. First, by setting $P = 2\beta$, the model of \citet{Johnson.1987} given in equation \eqref{eq:Johnson_model} is recovered. Thus, the model of \citet{Johnson.1987} has the inherent assumption
\begin{equation}
    \statictortuosity = \tortuosity \left(1+\frac{P}{2}\right)= \tortuosity \left(1+\frac{2\,\tortuosity\,\permeability}{\porosity\,\Lambda^2}\right)
    \label{eq:Johnson_implied_static_viscous_tortuosity}
\end{equation}
for the static viscous tortuosity. Second, the unsteady Darcy equation \eqref{eq:unsteady_Darcy_equation} can be recast as a dynamic permeability model by taking the Fourier transform:
\begin{equation}
	\FT{\superficialavg{\Vec{u}}}{\omega}  = -\frac{1}{\mu} \frac{\permeability}{1+\frac{\statictortuosity}{\tortuosity}\,\imag\frac{\omega}{\Omega_0}} \FT{\Vec{\nabla}\!\intrinsicavg{p}}{\omega} \,.
    \label{eq:unsteady_Darcy_frequency_domain}
\end{equation}
On the other hand, a Taylor expansion of the denominator of the model of \citet{Pride.1993} (equation \ref{eq:Pride_model}) at $\omega=0$ results in
\begin{equation}
\hat{\permeability}(-\omega) = \frac{\permeability}{1+\frac{\statictortuosity}{\tortuosity}\,\imag \frac{\omega}{\Omega_0}+ \mathit{O}\left(\frac{\omega^2}{\Omega_0^2}\right)} \,.
\end{equation}
By comparison, we see that the Darcy equation and the unsteady Darcy equation represents the zeroth and first order asymptotes to the low frequency behaviour of the frequency response function of the \citet{Pride.1993} model. This explains the observations of \citet{Zhu.2016a} that the unsteady Darcy equation with $\statictortuosity$ shows excellent agreement with the direct numerical simulations at low frequencies whereas a different time constant -- the high-frequency limit of the dynamic tortuosity $\tortuosity$ -- has to be employed at high frequencies. Notably, the high-frequency limit of the dynamic tortuosity $\tortuosity$ is a consistent choice of coefficient for the unsteady Darcy equation in that it leads to the correct limits for $\Omega\to 0$ and $\Omega \to \infty$.

In figure \ref{fig:dyn_permeability} the dynamic permeabilities implied by the different models are compared to the numerical simulations of \citet{Zhu.2016a} of oscillatory flow through a hexagonal sphere pack. It can be seen that the dynamic permeability models of \citet{Johnson.1987} and \citet{Pride.1993} show an excellent agreement with the simulation data. Note that for the hexagonal sphere pack, the assumption \eqref{eq:Johnson_implied_static_viscous_tortuosity} is fulfilled with an error of only $1 \%$, rendering the models of \citet{Johnson.1987} and \citet{Pride.1993} virtually identical for this geometry. On the other hand, the unsteady Darcy equation departs from the simulation data at medium or high frequencies, depending on the choice of the acceleration coefficient. The model of \citet{Turo.2013} will be discussed in the next section.

In conclusion, we find that the unsteady Darcy equation and the dynamic permeability model of \citet{Johnson.1987} can be seen as simplifications of the dynamic permeability model of \citet{Pride.1993}, which is able to accurately describe the simulation data for the hexagonal sphere pack.

\subsubsection{Improvement of the model of Turo \& Umnova (2013)}
\label{sec:improvement}

In the linear limit, the model \eqref{eq:Turo_model} corresponds to a dynamic permeability of the following form
\begin{equation}
    \hat{\permeability}(\Omega)= \frac{\permeability}{1  + \sqrt{ -\imag P \frac{\Omega}{\Omega_0}} - \imag \frac{\Omega}{\Omega_0}} \,.
    \label{eq:Turo_Umnova_frequency_domain}
\end{equation}
It can be seen that for $\Omega\to 0$ and for $\Omega\to\infty$ the correct limits \eqref{eq:Darcy_equation} and \eqref{eq:Johnson.1987_high_frequency_aymptotics} are approached. However, when the dynamic permeability is compared to the models of \citet{Johnson.1987,Pride.1993,Zhu.2014} it becomes apparent that the model of \citet{Turo.2013} severely underestimates the permeability at low and intermediate frequencies. This can be seen clearly in figure \ref{fig:dyn_permeability}a.

The analysis of this model deficiency suggests a simple remedy: We replace the underlying dynamic permeability model of equation \eqref{eq:Turo_model} with the model of \citet{Pride.1993} which has the correct behaviour at low frequencies. We obtain the following equation:
\begin{equation}
	\begin{split}
		\rho\,\frac{\mathrm{d}\!\superficialavg{\Vec{u}}}{\mathrm{d}t} =& -\frac{\porosity}{\tortuosity} \Vec{\nabla}\!\intrinsicavg{p} - \left[\frac{\porosity}{\tortuosity}\frac{\mu}{\permeability}\left(1+\xi \vert\!\superficialavg{\Vec{u}}\vert\right)-\frac{2\mu}{\Lambda^2\,\beta}\right] \superficialavg{\Vec{u}} 
		\\
		&-\frac{2 \rho \sqrt{\nu}}{\Lambda}\int_{-\infty}^{t} \left(\frac{\nu \superficialavg{\Vec{u}}}{\Lambda^2\,\beta^2}+\frac{\mathrm{d}\!\superficialavg{\Vec{u}}}{\mathrm{d}\tau}\right)\frac{\euler^{- \frac{\nu (t-\tau)}{\Lambda^2\,\beta^2}  }}{\sqrt{\pi(t-\tau)}}\,\mathrm{d}\tau \,.
	\end{split}
    \label{eq:extended_Pride_et_al_model}
\end{equation}
This model allows us to explore the potential of the basic idea of \citet{Turo.2013} of combining a dynamic permeability model with the Forchheimer nonlinearity.

\subsubsection{Lower bounds for the coefficients of the unsteady Forchheimer equation}
\label{sec:implied_energy_equation}

In this section, we discuss lower bounds for the coefficients of the unsteady Forchheimer equation that arise from Kelvin's minimum energy theorem, Helmholtz' minimum dissipation theorem and the volume-averaged kinetic energy equation if the coefficients are considered as constants.

First, we multiply the unsteady Forchheimer equation \eqref{eq:unsteady_Darcy-Forchheimer_equation} with the superficial velocity such that after some rearrangements the following equation is obtained:
\begin{equation}
    \frac{\mathrm{d}}{\mathrm{d}t}\left(\frac{c}{2}\superficialavg{\Vec{u}}^2\right) =-\superficialavg{\Vec{u}}\cdot\Vec{\nabla}\!\intrinsicavg{p}-a \superficialavg{\Vec{u}}^2- b\left\vert\superficialavg{\Vec{u}}\right\vert \superficialavg{\Vec{u}}^2
    \label{eq:implied_energy_equation}
\end{equation}
where the first term on the right hand side is the power per unit volume added to or removed from the flow by the macroscopic pressure gradient. Comparing this equation to the volume-averaged kinetic energy equation \citep{Zhu.2014}
\begin{equation}
    \frac{\mathrm{d}}{\mathrm{d}t}\left(\frac{1}{2}\rho\superficialavg{\Vec{u}^2}\right) = -\superficialavg{\Vec{u}}\cdot\Vec{\nabla}\!\intrinsicavg{p} - 2\mu\superficialavg{\Tensor{S}:\Tensor{S}}\,,
\end{equation}
wherein $\Tensor{S}$ is the strain rate tensor, we find that the term on the left hand side of \eqref{eq:implied_energy_equation} is the time derivative of a positive quantity. Hence, it can take both signs and may be arbitrarily large. Therefore, it cannot be part of the dissipation. On the other hand, if parts of the second and third terms on the right hand side of \eqref{eq:implied_energy_equation} belonged to the time derivative of the kinetic energy, then for a steady flow the kinetic energy would increase with time, which contradicts the assumption of a steady flow. Consequently, we identify
\begin{subequations}
    \begin{align}
        \frac{1}{2}\rho\superficialavg{\Vec{u}^2} &= \frac{c}{2} \superficialavg{\Vec{u}}^2 \\
        2\mu\superficialavg{\Tensor{S}:\Tensor{S}} &= \left(a + b\left\vert\superficialavg{\Vec{u}}\right\vert\right)\superficialavg{\Vec{u}}^2
    \end{align}
\end{subequations}
as underlying assumptions of the unsteady Forchheimer equation. Note that the dissipation term is consistent with \citet{Nield.2000} who discussed stationary flow.

From Kelvin's minimum energy theorem \citep[p.384]{Batchelor.2000}, it is possible to show that the kinetic energy for a given superficial velocity is smallest in the potential flow. This results in the inequality \citep[p.123]{Lafarge.1993}
\begin{equation}
    \frac{1}{2}\rho\superficialavg{\Vec{u}^2} \ge \frac{1}{2} \frac{\rho\,\tortuosity}{\porosity}\superficialavg{\Vec{u}}^2
\end{equation}
which is valid for an isotropic porous medium. Moreover, according to Helmholtz' minimum dissipation theorem \citep[pp.227--228]{Batchelor.2000}, the dissipation rate for a given superficial velocity is smallest in the Stokes flow. Using the expression of \citet{Zhu.2014,Paez-Garcia.2017} for the dissipation rate in the Stokes flow, we arrive at the inequality
\begin{equation}
    2\mu\superficialavg{\Tensor{S}:\Tensor{S}} \ge \frac{\mu}{\permeability} \superficialavg{\Vec{u}}^2
\end{equation}
which again is valid for an isotropic porous medium. We therefore obtain the lower bounds
\begin{subequations}
    \begin{align}
        a &\ge \frac{\mu}{\permeability} \\
        b &\ge 0 \\
        c &\ge \frac{\rho\,\tortuosity}{\porosity}
    \end{align}
\end{subequations}
for the coefficients of the unsteady Forchheimer equation.

These inequalities could be interpreted as \textit{realisability conditions} similar to those of \citet{Schumann.1977} in the context of Reynolds stress turbulence models, for if these conditions are violated, no pore scale velocity field can be found such that equation \eqref{eq:implied_energy_equation} describes the evolution of the kinetic energy.

\section{Methodology}

In this section, we describe the simulation dataset that will be used as a reference for comparing the different models. Moreover, we discuss aspects of the numerical implementation of the models. Finally, we introduce the metrics that will be used for quantifying the model errors.

\subsection{Description of the flow solver}

The reference simulations were conducted using our in-house code MGLET \citep{Manhart.2001} which solves the incompressible Navier-Stokes equations on a Cartesian block-structured grid with a staggered arrangement of variables. For the spatial discretisation, a symmetry-preserving second-order finite volume scheme \citep{Verstappen.2003} is employed. The no-slip boundary condition on the spheres is imposed by means of a mass-conserving ghost-cell immersed boundary method \citep{Peller.2006,Peller.2010}. For the temporal discretisation, a third-order explicit Runge-Kutta method \citep{Williamson.1980} is used and a projection step \citep{Chorin.1968} is performed at every stage.

\subsection{Description of the simulation database}
\label{sec:description_of_the_simulation_dataset}

The simulation database that will be used as a reference for the model comparison consists of direct numerical simulations of oscillatory flow through a hexagonal sphere pack \citep{Unglehrt.2022,Unglehrt.2023,Unglehrt.2023_DLES}. The sphere pack is a triply periodic close-packed arrangement of equal spheres of diameter $d$. Hence, each sphere is in contact with 12 other spheres and the porosity is given as $\porosity=1-\frac{\pi}{3\sqrt{2}}\approx 0.26$. The flow was driven by a sinusoidally time-varying pressure gradient
\begin{equation}
    \Vec{\nabla}\!\intrinsicavg{p}(t) = g_x \,\sin(\Omega t)
    \label{eq:macroscopic_pressure_gradient}
\end{equation}
and therefore depends on two dimensionless parameters: The Hagen number $\Hg=\left\vert g_x\right\vert d^3/(\rho \nu^2)$ determines the amplitude and the Womersley number $\Wo=\sqrt{\Omega\,d^2/\nu}$ determines the frequency of the macroscopic pressure gradient. Geometrically, the Hagen number can be seen as the ratio of the distance over which the pressure gradient can move a particle within a viscous time unit $d^2/\nu$ to the sphere diameter $d$. Meanwhile, the Womersley number can be seen as the ratio of the sphere diameter to the thickness of the oscillatory boundary layer $\sqrt{\nu/\Omega}$. The Reynolds number was defined as 
\begin{equation}
    \Rey = \limsup\limits_{t\to\infty} \frac{\left\vert\superficialavg{\Vec{u}}\right\vert d}{\nu}
\end{equation}
and was obtained as a result of the simulations. The simulations were performed at the three Womersley numbers $\Wo=10$, $31.62$, and $100$ that correspond to the low, medium and high frequency regime in linear flow, respectively.\footnote{
    The medium frequency regime in linear flow through a hexagonal sphere pack occurs around a Womersley number $\Wo_0 =\sqrt{\Omega_0 d^2/\nu}= \sqrt{\porosity\,d^2/(\tortuosity\,\permeability)}= 30.5$, where $\Omega_0= \porosity\,\nu/(\tortuosity\,\permeability)$ is the transition frequency \citep{Pride.1993}. The geometric constants are given in table \ref{tab:geometric_constants}.
} The Hagen numbers were set such that the simulations cover linear flow, laminar nonlinear flow, and transitional and turbulence-like flow at each Womersley number. The flow was started from rest and simulated until a recurrent behaviour could be observed in the superficial velocity. The difference in the superficial velocity in the last simulated cycle and the preceding cycle was less than approximately $0.1\%$ of the maximum superficial velocity in the last simulated cycle for the laminar cases and about $2\%$ for the transitional and turbulence-like cases. The simulation parameters can be found in table \ref{tab:simulation_parameters}.

\begin{table}[t]
\begin{center}
\begin{minipage}{\textwidth}
\caption{Parameters of the simulations of oscillatory flow through a hexagonal sphere pack.}\label{tab:simulation_parameters}%
\begin{tabular}{@{}llllll@{}}
\toprule
Case & $\Hg$  & $\Wo$ & $\Rey$ & number of simulated cycles & flow regime \\
\midrule
LF1 \footnotemark[1] & $1.00 \cdot 10^{3}$ & $10$ & $0.171$ & $1.5$ & linear \\
LF2 \footnotemark[1] & $1.00 \cdot 10^{4}$ & $10$ & $1.7$ & $2.25$ & linear \\
LF3 \footnotemark[1] & $1.00 \cdot 10^{5}$ & $10$ & $14.8$ & $1.4$ & laminar nonlinear \\
LF4 \footnotemark[1] & $1.00 \cdot 10^{6}$ & $10$ & $76.7$ & $1.25$ & laminar nonlinear \\
LF5 \footnotemark[2] & $3.16 \cdot 10^{6}$ & $10$ & $158$ & $2.27$ & transitional \\
LF6 \footnotemark[2] & $1.00 \cdot 10^{7}$ & $10$ & $307$ & $1.56$ & turbulence-like \\
\midrule
MF1 \footnotemark[1] & $1.00 \cdot 10^{4}$ & $31.6$ & $0.857$ & $3$ & linear \\
MF2 \footnotemark[1] & $1.00 \cdot 10^{5}$ & $31.6$ & $8.57$ & $3$ & linear \\
MF3 \footnotemark[1] & $3.16 \cdot 10^{5}$ & $31.6$ & $26.9$ & $3$ & laminar nonlinear \\
MF4 \footnotemark[1] & $1.00 \cdot 10^{6}$ & $31.6$ & $73.1$ & $3$ & laminar nonlinear \\
MF5 \footnotemark[2] & $3.16 \cdot 10^{6}$ & $31.6$ & $157$ & $6.4$ & transitional \\
MF6 \footnotemark[2] & $1.00 \cdot 10^{7}$ & $31.6$ & $298$ & $2.26$ & turbulence-like \\
\midrule
HF1 \footnotemark[1] & $1.00 \cdot 10^{5}$ & $100$ & $1.3$ & $20.4$ & linear \\
HF2 \footnotemark[1] & $1.00 \cdot 10^{6}$ & $100$ & $13$ & $19.9$ & linear \\
HF3 \footnotemark[1] & $1.00 \cdot 10^{7}$ & $100$ & $132$ & $6.32$ & laminar nonlinear \\
HF4 \footnotemark[1] & $1.78 \cdot 10^{7}$ & $100$ & $252$ & $8$ & laminar nonlinear \\
HF5 \footnotemark[2] & $3.16 \cdot 10^{7}$ & $100$ & $465$ & $6$ & transitional \\
HF6 \footnotemark[3] & $1.00 \cdot 10^{8}$ & $100$ & $1090$ & $3.91$ & turbulence-like \\
HF7 \footnotemark[3] & $1.00 \cdot 10^{9}$ & $100$ & $3620$ & $10$ & turbulence-like \\
\botrule
\end{tabular}
\footnotetext[1]{from \citet{Unglehrt.2022}}
\footnotetext[2]{from \citet{Unglehrt.2023}}
\footnotetext[3]{from \citet{Unglehrt.2023_DLES}}
\end{minipage}
\end{center}
\end{table}

\begin{figure}
    \centering
    \includegraphics[width=\textwidth,axisratio=1.618]{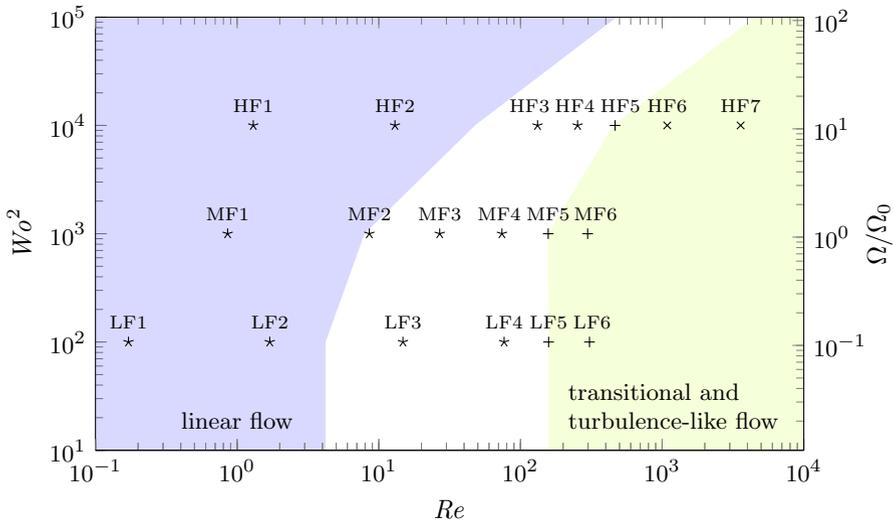}
    \caption{Simulation parameters. The stars mark the simulations from \citep{Unglehrt.2022}, the pluses mark the simulations from \citep{Unglehrt.2023} and the crosses mark the simulations from \citep{Unglehrt.2023_DLES}. The regions shaded in blue and green corresponds to the linear regime and the transitional and turbulence-like regime, respectively.}
    \label{fig:simulation_parameters}
\end{figure}

Figure \ref{fig:simulation_parameters} shows the dimensionless parameters of our simulation database in the $\Rey$--$\Wo^2$ parameter space. We additionally give the frequency ratio $\Omega/\Omega_0$ where $\Omega_0=930.25\,\nu/d^2$ is the transition frequency defined in section \ref{sec:dynamic_permeability_models} that can be used to distinguish between the low and the high frequency regime in linear flow. The parameter region for linear flow was defined such that the first harmonic contains $99 \%$ of the energy of the velocity field \citep{Unglehrt.2022}. The breaking of the geometry-imposed symmetries in the flow was used to distinguish the laminar nonlinear from the transitional and turbulence-like regime \citep{Unglehrt.2022_TSFP}.

\begin{figure}
    \centering
    \includegraphics[width=\textwidth]{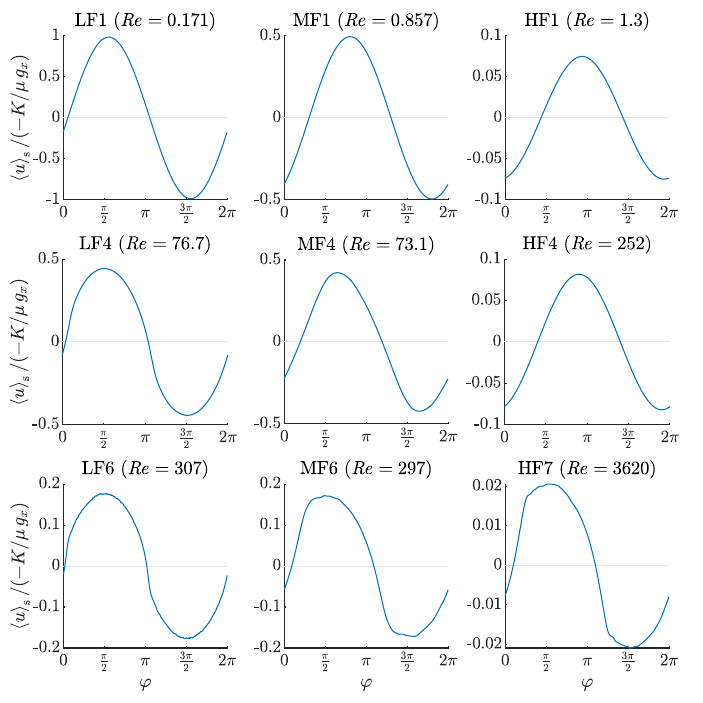}
    \caption{Variation of the superficial velocity over the cycle from the direct numerical simulations of oscillatory flow through a hexagonal sphere pack. In all cases, the forcing is given by $\Vec{\nabla}\!\intrinsicavg{p} = g_x \,\sin\varphi$ and the superficial velocity is normalised by the value obtained from Darcy's law at the peak pressure gradient.}
    \label{fig:superficial_velocity}
\end{figure}

The time series of the superficial velocity for some of the simulation cases is shown in figure \ref{fig:superficial_velocity}. The amplitudes of the superficial velocity decrease with increasing Reynolds or Womersley number. The decrease of the amplitude with Womersley number is caused by the inertia of the system, whereas the decrease with Reynolds number can be explained by the increase of the drag. As the Reynolds number increases, the behaviour of the superficial velocity becomes non-sinusoidal. This could be explained mainly by the effect of nonlinearity \citep{Unglehrt.2023}. Turbulence becomes significant only for the cases HF6 and HF7 \citep{Unglehrt.2023_DLES}. At high Womersley and Reynolds numbers, we can distinguish phases of strong and weak acceleration that lead to distinctive bends in the superficial velocity. In \citep{Unglehrt.2023} we could link this to an increase in the convective pressure drag due to flow separation at the contact points. As the Womersley number increases, we observe a phase lag of the superficial velocity with respect to the forcing. This can be seen from the zero-crossings of the superficial velocity which do not coincide with the zeros of the forcing ($\varphi = k \pi$ for $k=0,1,2,\dots$). This is due to the increasing importance of the inertia compared to the drag. On the other hand, the phase lag decreases as the Reynolds number increases. This goes in hand with an increase of the nonlinear drag which advances the inertial term. Overall, these time series demonstrate a distinct effect of nonlinearity and turbulence on the response of the superficial velocity to harmonic excitations.

\subsection{Model constants for the hexagonal sphere pack}

In this section, we specify the model constants that were used to evaluate the model predictions. The parameters for the linear models have rigorous definitions in terms of the steady Stokes flow ($\permeability$, $\statictortuosity$) and the potential flow solution ($\Lambda/d$, $\tortuosity$), respectively. Notably, these quantities are intrinsic properties of the pore geometry. The low frequency properties of the hexagonal sphere pack were determined from direct numerical simulations in the studies of \citet{Zhu.2016a} and \citet{Sakai.2020}; the high frequency properties were determined by a potential flow calculation in our previous study \citep{Unglehrt.2023}. The parameter values are summarised in table \ref{tab:geometric_constants}. The permeability value $1.731\cdot 10^{-4}\,d^2$ agrees well with the Kozeny-Carman correlation ($1.771\cdot 10^{-4}\,d^2$) and the self-consistent estimate of \citet{Boutin.2010} ($\permeability_c=1.769\cdot 10^{-4}\,d^2$).

For the coefficients of the unsteady Forchheimer equation various choices can be found in the literature (see section \ref{sec:unsteady_Forchheimer_equation}). Following \citet{Sollitt.1972}, we assume that the coefficients $a$ and $b$ are constant and take their steady state values. This choice ensures that the unsteady Forchheimer equation has the correct low frequency limit behaviour. For the acceleration coefficient $c$, we consider two choices: Either the coefficient is set to the potential flow value $\rho\,\tortuosity/\porosity$ such that the correct high-frequency behaviour is recovered \citep{Zhu.2016a}, or, following \citet{Zhu.2016}, the coefficient is set to the Stokes flow value $\rho\,\statictortuosity/\porosity$ such that the unsteady Forchheimer equation reduces to the unsteady Darcy equation with the static viscous tortuosity \citep{Zhu.2014}. We do not consider time- or frequency-dependent coefficients as no generally applicable correlations have been given in the literature. Note that the coefficients of the unsteady Forchheimer equation could be adjusted to improve the fit for some simulation cases; however, this comes at the price of high prediction errors behaviour in some of the asymptotic limits, e.g. for slowly varying flow or for linear flow.

\begin{figure}
    \centering
    \includegraphics[width=0.5\textwidth]{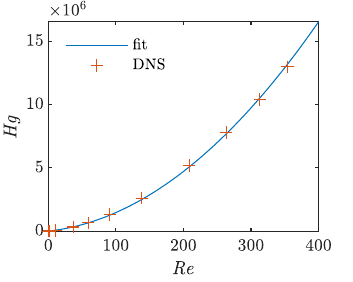}
    \caption{Forchheimer equation \eqref{eq:Forchheimer_equation} with coefficients \eqref{eq:Forchheimer_coefficients_fit} and direct numerical simulation data for stationary flow through a hexagonal sphere pack.}
    \label{fig:Forchheimer_fit}
\end{figure}

The coefficients of the Forchheimer equation \eqref{eq:Forchheimer_equation} are determined based on the simulation results in \citep{Sakai.2020,Unglehrt.2023}. The linear Forchheimer coefficient $a$ is chosen such that the Forchheimer equation approaches Darcy's law for $\Rey\to 0$. Then, the nonlinear Forchheimer coefficient $b$ is determined from a least squares fit to the ratio of the macroscopic pressure gradient and the superficial velocity. We obtained the following coefficients:
\begin{subequations}
    \begin{align}
        a &= \frac{\mu}{\permeability}=5777 \,\frac{\mu}{d^2} \\
        b &= 88.9\,\frac{\rho}{d} \,.
    \end{align}
    \label{eq:Forchheimer_coefficients_fit}
\end{subequations}
It can be seen in figure \ref{fig:Forchheimer_fit} that these coefficients provide a good fit to the simulation data. The values also agree well with the Ergun equation for packed beds \citep{Macdonald.1979}, which predicts $a=5647\,\mu/d^2$ and $b=76.3\,\rho/d$. Finally, the nonlinear coefficient $\xi$ in the equations \eqref{eq:Turo_model} and \eqref{eq:extended_Pride_et_al_model} was calculated as $b \permeability/\mu=0.0154\,d/\nu$.

\begin{table}[t]
\begin{center}
\begin{minipage}{\textwidth}
    \caption{Geometric parameters for the hexagonal close-packed arrangement of equal spheres.}
    \label{tab:geometric_constants}%
    \begin{tabular}{@{}lll@{}}
        \toprule
        Parameter & Symbol & Value \\
        \midrule
        Porosity & $\porosity$ & $1-\frac{\pi}{3\sqrt{2}}=0.2595$\\
        Permeability & $\permeability$ & $ 1.731\cdot 10^{-4}\,d^2$ \footnotemark[1] \\
        Low-frequency limit of the dynamic tortuosity & $\statictortuosity$ & $2.657$ \footnotemark[2] \\
        High-frequency limit of the dynamic tortuosity & $\tortuosity$ & $1.622$ \footnotemark[3] \\
        Boundary layer length scale & $\Lambda$ & $5.904\cdot10^{-2}\,d$ \footnotemark[3] \\
        \botrule
    \end{tabular}
    \footnotetext[1]{from \citet{Sakai.2020}}
    \footnotetext[2]{from \citet{Zhu.2016a}}
    \footnotetext[3]{from \citet{Unglehrt.2023}}
\end{minipage}
\end{center}
\end{table}

\subsection{Implementation of the models}

In this section, we describe and verify the discretisation schemes that were used to evaluate the model predictions. The full derivation of these discretisations is given in \citep[appendix B.2]{Unglehrt.2024}.

\subsubsection{Discretisation of the model of Pride et al. (1993)}

As we would like to obtain model predictions for transient flow, we discretise the model of \citet{Pride.1993} in the time domain. We discretise equation \eqref{eq:Pride_model_time_domain} using the implicit Euler method and a piecewise linear interpolation of the convolution term. We obtained the following scheme
\begin{subequations}
	\begin{equation}
        a \superficialavg{\Vec{u}}^{n+1} =  \superficialavg{\Vec{u}}^{n} - {\Delta t}\,\frac{\porosity}{\rho\,\tortuosity}\Vec{\nabla}\!\intrinsicavg{p}^{n+1} -\sum_{k=1}^{\infty} c_k \superficialavg{\Vec{u}}^{n-k+1}
	\end{equation}
	with precomputable coefficients
	\begin{align}
    	a =& 1 + h\,\left(\frac{4\beta^2}{P}-2\beta\right) + 2\beta \left[\left(\frac{1}{2}+h\right) \mathrm{erf}\!\left(\sqrt{h}\right)  + \sqrt{\frac{h}{\pi}}\euler^{-h}\right]\\
    	\begin{split}
        	c_k =&  -2\beta \left\lbrace \left[\mathrm{erf}\!\left(\sqrt{\xi}\right)\right]_{ (k-1) h}^{ k h} \left(\frac{1}{2}+(k-1) h\right) + \left[\sqrt{\frac{\xi}{\pi}}\euler^{-\xi}\right]_{ (k-1) h}^{ k h}\right\rbrace \\
        	&+ 2\beta \left\lbrace  \left[\mathrm{erf}\!\left(\sqrt{\xi}\right)\right]_{ k h}^{ (k+1) h} \left(\frac{1}{2}+(k+1)h\right) + \left[\sqrt{\frac{\xi}{\pi}}\euler^{-\xi}\right]_{ k h}^{ (k+1) h}\right\rbrace
    	\end{split}
	\end{align}
	\label{eq:Pride_model_discretisation}
	depending on the dimensionless time step $h = \frac{\nu {\Delta t}}{\Lambda^2 \beta^2}$.
\end{subequations}
Note that the computational effort grows linearly with the number of time steps as the discrete convolution must be performed over a time series of increasing length. Computationally more efficient discretisations could be devised by approximating the tail of the convolution kernel in equation \eqref{eq:Pride_model_time_domain} with an exponential function similar to \citep{vanHinsberg.2011}.

The correctness of the scheme \eqref{eq:Pride_model_discretisation} and its convergence with the time step size is assessed using the method of manufactured solutions. As a test case, we consider the velocity $\superficialavg{u} = j_0\,\frac{t^2}{2} \, \theta(t)$ with the Heaviside function $\theta(t)$. The forcing that is necessary to find this velocity as a solution to \eqref{eq:Pride_model_time_domain} can be obtained as
\begin{equation}
    \begin{split}
        -\Vec{\nabla}\!\intrinsicavg{p} =& \frac{\rho\,\tortuosity}{\porosity} j_0\,t\, \theta(t) + \left(\frac{\mu}{\permeability}-\frac{\tortuosity}{\porosity} \frac{2\mu}{\Lambda^2\,\beta}\right) j_0 \,\frac{t^2}{2} \, \theta(t)  \\
        &+\frac{ \rho\,\tortuosity}{\porosity}\,\beta\,j_0\,\theta(t)\,\left[\left(t + \frac{\nu t^2}{\Lambda^2 \beta^2}-\frac{1}{4}\frac{\Lambda^2\,\beta^2}{\nu}\right)\mathrm{erf}\!\left(\sqrt{\frac{\nu t}{\Lambda^2 \beta^2}}\right) \right.\\
        &\left.+\left(t +\frac{\Lambda^2\,\beta^2}{2 \nu} \right) \sqrt{\frac{\nu t}{\Lambda^2 \beta^2}} \frac{ \euler^{-\frac{\nu t}{\Lambda^2 \beta^2}}}{\sqrt{\pi}}\right] \,.
    \end{split}
    \label{eq:verification_test_forcing__Pride_model}
\end{equation}

\begin{figure}
	\centering
    \subcaptionbox{\label{fig:verification_test_forcing__Pride_model}}{
        \includegraphics[width=0.45\textwidth]{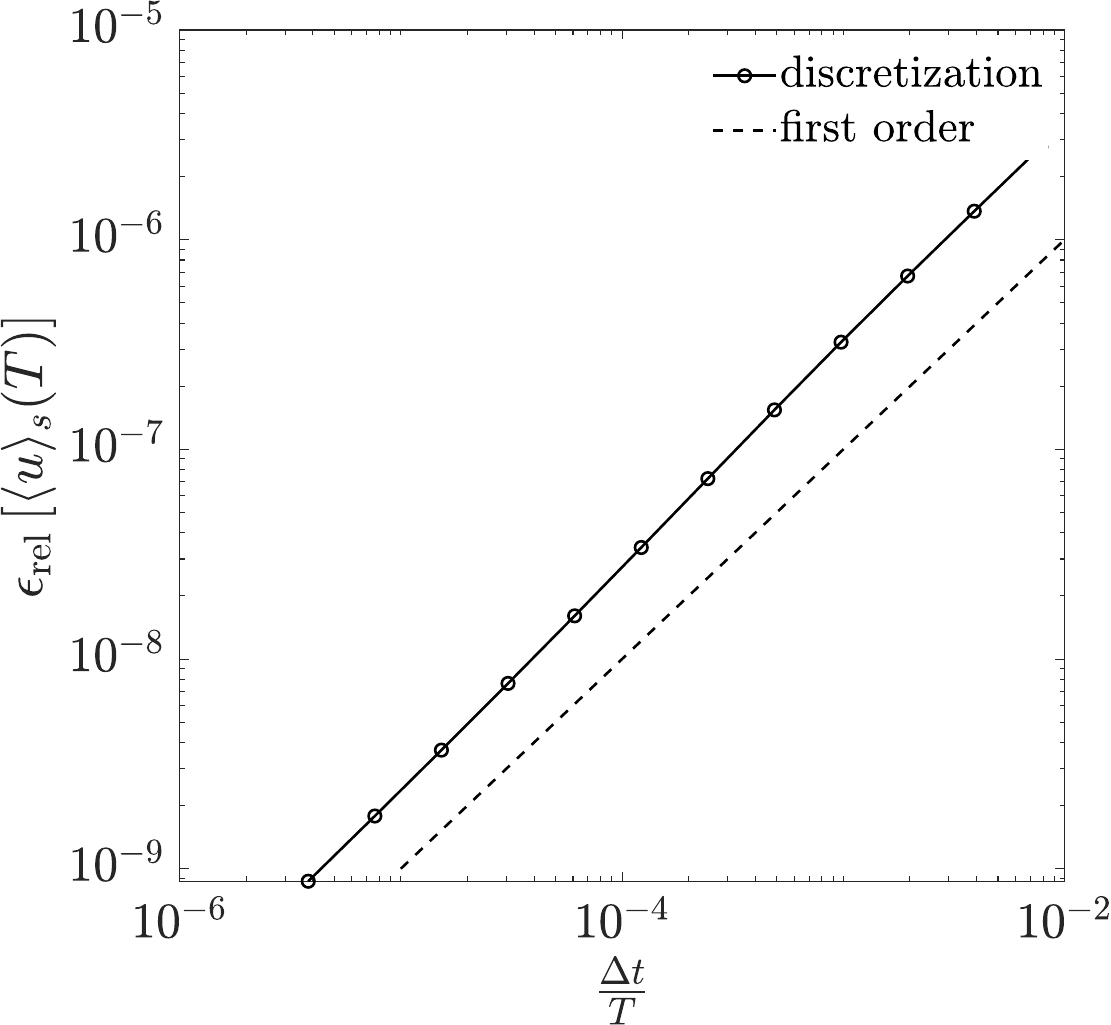}
    }
    \subcaptionbox{\label{fig:verification_test_forcing__Turo_model}}{
    	\includegraphics[width=0.45\textwidth]{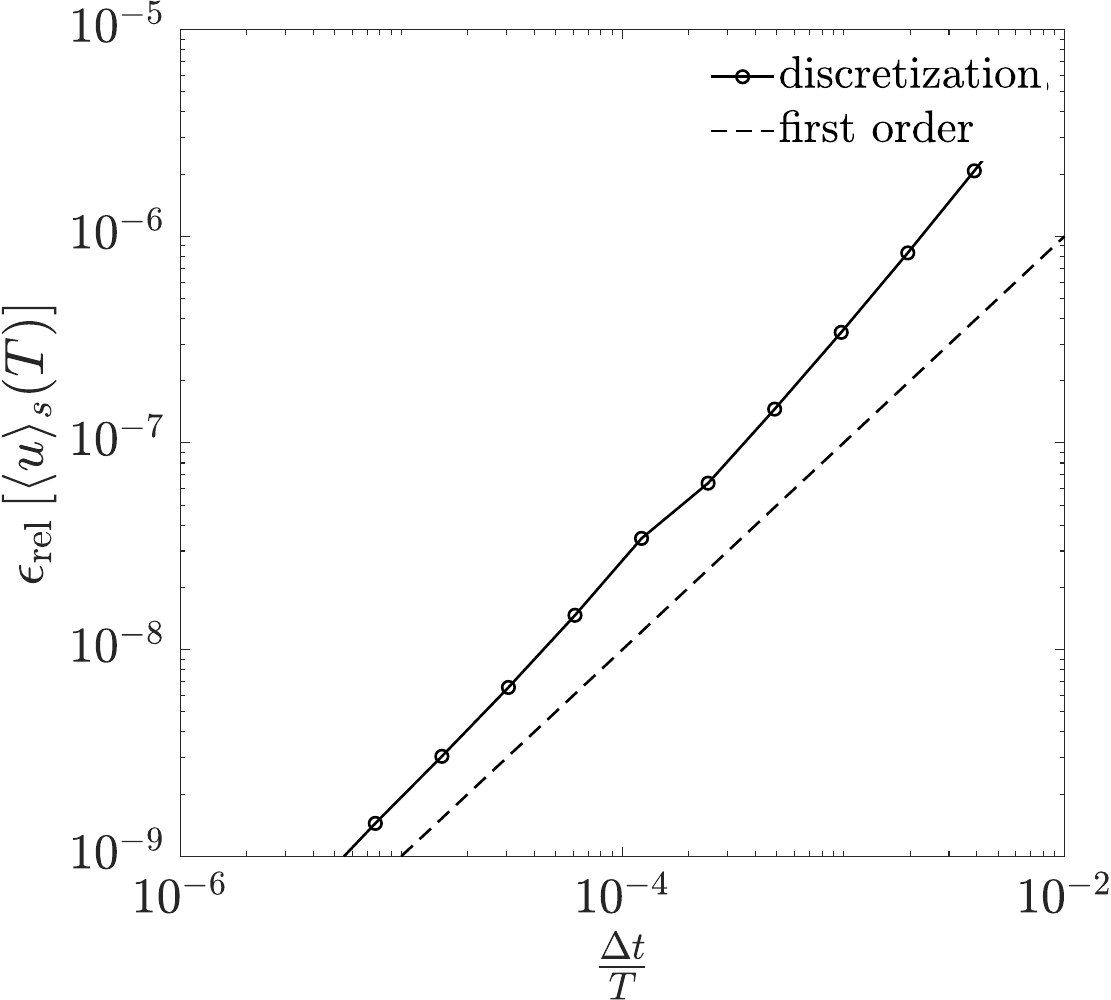}
    }
	\caption{Verification of the discretisations for (a) the model of \citet{Pride.1993} and (b) the model of \citet{Turo.2013} with manufactured solutions (equation \eqref{eq:verification_test_forcing__Pride_model} and equation \eqref{eq:verification_test_forcing__Turo_model} with $\frac{\xi }{j_0^{1/5} \nu^{2/5}}=0.01$, respectively).}
\end{figure}

We simulated \eqref{eq:Pride_model_discretisation} with the forcing defined above for a hexagonal sphere pack. Figure \ref{fig:verification_test_forcing__Pride_model} shows the relative error of $\superficialavg{u}$ with respect to the analytic solution at $\frac{\nu T}{d^2}=5$. We observe first order convergence which means that the singularity in the integral kernel is treated with sufficient accuracy.

\subsubsection{Discretisation of the model of Turo \& Umnova (2013)}

Similar to the preceding section, the discretisation of equation \eqref{eq:Turo_model} was derived using the implicit Euler method and a linear interpolation in the convolution term. This resulted in the following scheme
\begin{subequations}
	\begin{equation}
	a \superficialavg{\Vec{u}}^{n+1} + b \left\vert \superficialavg{\Vec{u}}^{n+1}\right\vert \superficialavg{\Vec{u}}^{n+1} =  \superficialavg{\Vec{u}}^{n} - {\Delta t}\,\frac{\porosity}{\rho\,\tortuosity}\Vec{\nabla}\!\intrinsicavg{p}^{n+1} -\sum_{k=1}^{\infty} c_k \superficialavg{\Vec{u}}^{n-k+1}
	\end{equation}
	with precomputable coefficients
	\begin{align}
	a =& 1 + \frac{\porosity}{\tortuosity}\frac{\nu \Delta t}{\permeability} + \frac{2\sqrt{\nu  \Delta t}}{\Lambda}\frac{2}{\sqrt{\pi}}    \\
	b =& \frac{\porosity}{\tortuosity}\frac{\nu \Delta t}{\permeability} \xi \\
	c_k =&  \frac{2\sqrt{\nu \Delta t}}{\Lambda}\frac{2}{\sqrt{\pi}} \left(\sqrt{k+1}- 2\sqrt{k}+\sqrt{k-1}\right)  \,.
	\end{align}
	\label{eq:Turo_model_discretisation}
\end{subequations}
The nonlinear equation appearing in every time step was solved with the function \texttt{fsolve} in MATLAB where the superficial velocity from the previous time step was taken as an initial guess. Other discretisations of the fractional derivative have been given e.g. by \citet{Diethelm.2005}. 

As in the preceding section, we verified the correctness and convergence of our implementation using a manufactured solution for the velocity $\superficialavg{u} = j_0\,\frac{t^2}{2} \, \theta(t)$. The corresponding forcing results from \eqref{eq:Turo_model} as
\begin{equation}
    \begin{split}
        - \Vec{\nabla}\!\intrinsicavg{p} = & \frac{\rho \tortuosity}{\porosity}  j_0\,t\, \theta(t) + \frac{\mu}{\permeability} \left(1+\xi\,\vert j_0\vert\,\frac{t^2}{2} \right) j_0\,\frac{t^2}{2} \, \theta(t) \\
        &+\frac{\rho \tortuosity}{\porosity} j_0\,\frac{\sqrt{\nu}}{\Lambda}  \frac{8}{3\sqrt{\pi}} t^{\frac{3}{2}} \theta(t) \,.
    \end{split}
    \label{eq:verification_test_forcing__Turo_model}
\end{equation}
Figure \ref{fig:verification_test_forcing__Turo_model} shows the convergence of the numerical solution to the analytical solution at a time $\frac{\nu T}{d^2}=5$ and for a coefficient of nonlinearity $\frac{\xi }{j_0^{1/5} \nu^{2/5}}=0.01$. Again, we have used the material properties of the hexagonal sphere pack.

\subsubsection{Evaluation of model ODEs}

The unsteady Darcy equation \eqref{eq:unsteady_Darcy_equation} and the unsteady Forchheimer equation \eqref{eq:unsteady_Darcy-Forchheimer_equation} are simple ordinary differential equations. These were solved using a second order modified Rosenbrock method implemented in the MATLAB routine \texttt{ode23s} \citep{Shampine.1997}. The time step size was chosen adaptively according to a relative tolerance of $10^{-6}$ and an absolute tolerance of $10^{-8}\,\frac{\nu}{d}$ for the superficial velocity.

\subsection{Metrics for comparison}

In order to compare the different models between the different simulation cases, we define an integral error metric. For two periodic signals $u_1(t)$ and $u_2(t)$ that will represent the model predictions and the reference for the superficial velocity, respectively, the $L^2$ distance between the signals
\begin{equation}
    \left\Vert u_1 - u_2 \right\Vert_2 = \left(\frac{1}{T}\int_{T_{\mathrm{sim}}-T}^{T_{\mathrm{sim}}} \left\vert u_1 - u_2\right\vert^2 \,\mathrm{d}t \right)^\frac{1}{2} \,.
    \label{eq:L2_distance}
\end{equation}
As we are dealing with periodic signals, we can take advantage of their Fourier series to further split this error into an amplitude and a phase contribution. Using Parseval's identity, we can express the squared $L^2$ distance as
\begin{equation}
     \left\Vert u_1 - u_2 \right\Vert_2^2 = \sum_{k=-\infty}^{\infty} \left\vert c_k\right\vert^2
\end{equation}
where $c_k$ are the Fourier series coefficients of $u_1-u_2$
\begin{equation}
    c_k = \frac{1}{T}\int_{T_{\mathrm{sim}}-T}^{T_{\mathrm{sim}}} (u_1 - u_2) \euler^{-\imag k\frac{2\pi}{T} t} \,\mathrm{d}t = a_k - b_k
\end{equation}
where $a_k$ and $b_k$ are the Fourier series coefficients of $u_1$ and $u_2$, respectively. Expressing the complex numbers $a_k$ and $b_k$ in polar form, we obtain
\begin{equation}
    \begin{split}
        \left\Vert u_1 - u_2 \right\Vert_2^2 &= \sum_{k=-\infty}^{\infty} \left\vert \left\vert a_k\right\vert \euler^{\imag \phi_k}-\left\vert b_k\right\vert  \euler^{\imag \psi_k}\right\vert^2 \\
        &= \sum_{k=-\infty}^{\infty} \left\vert\left\vert a_k\right\vert - \left\vert b_k\right\vert\right\vert^2 + \sum_{k=-\infty}^{\infty} 2 \left\vert a_k\right\vert  \left\vert b_k\right\vert \left[1- \cos(\phi_k-\psi_k)\right] \,.
    \end{split}
\end{equation}
The first term represents the difference in the magnitude of the Fourier coefficients and the second term represents the difference in the phase of the Fourier coefficients. Since both terms are positive, we can define an amplitude and a phase distance
\begin{align}
    d_{\mathrm{ampl.}}(u_1,u_2) &= \sqrt{\sum_{k=-\infty}^{\infty} \left\vert\left\vert a_k\right\vert - \left\vert b_k\right\vert\right\vert^2} \label{eq:amplitude_distance} \\
    d_{\mathrm{phase}}(u_1,u_2) &= \sqrt{\sum_{k=-\infty}^{\infty} 2 \left\vert a_k\right\vert  \left\vert b_k\right\vert \left[1- \cos(\phi_k-\psi_k)\right]} \label{eq:phase_distance}
\end{align}
and the $L^2$ distance can be decomposed into the sum of squares \begin{equation}
     \left\Vert u_1 - u_2 \right\Vert_2^2=d_{\mathrm{ampl.}}(u_1,u_2)^2+d_{\mathrm{phase}}(u_1,u_2)^2 \,.
\end{equation}
Note that $d_{\mathrm{ampl.}}(u_1,u_2)$ and $d_{\mathrm{phase}}(u_1,u_2)$ are not proper distances in that they can be zero also if $u_1\neq u_2$. Moreover, $d_{\mathrm{phase}}(u_1,u_2)$ does not satisfy the triangle inequality. However, these definitions ensure that two identical waveforms that are shifted with respect to each other only lead to a phase distance, but not to an amplitude distance. Also, a waveform that is a constant multiple of the other only leads to an amplitude distance, but not to a phase distance.

\section{Results}

\subsection{Comparison of model errors}

In this section, we compare the accuracy of the model predictions in response to the forcing \eqref{eq:macroscopic_pressure_gradient} with respect to the direct numerical simulation dataset. In particular, we look at the model of \citet{Turo.2013} given in equation \eqref{eq:Turo_model}, the extended \citet{Pride.1993} model given in equation \eqref{eq:extended_Pride_et_al_model} and the unsteady Forchheimer equation \eqref{eq:unsteady_Darcy-Forchheimer_equation}. For the unsteady Forchheimer equations, we take the acceleration coefficient as the static viscous tortuosity $\statictortuosity$ and as the high-frequency limit of the dynamic tortuosity $\tortuosity$. For reference, we also include the unsteady Darcy equation \eqref{eq:unsteady_Darcy_equation} and the models of \citet{Johnson.1987} and \citet{Pride.1993} given by equation \eqref{eq:Pride_model_time_domain}, where for the model of \citet{Johnson.1987} the relation \eqref{eq:Johnson_implied_static_viscous_tortuosity} was used.

\begin{table}
	\centering
    \subcaptionbox{}{
        \includegraphics[width=0.9\textwidth]{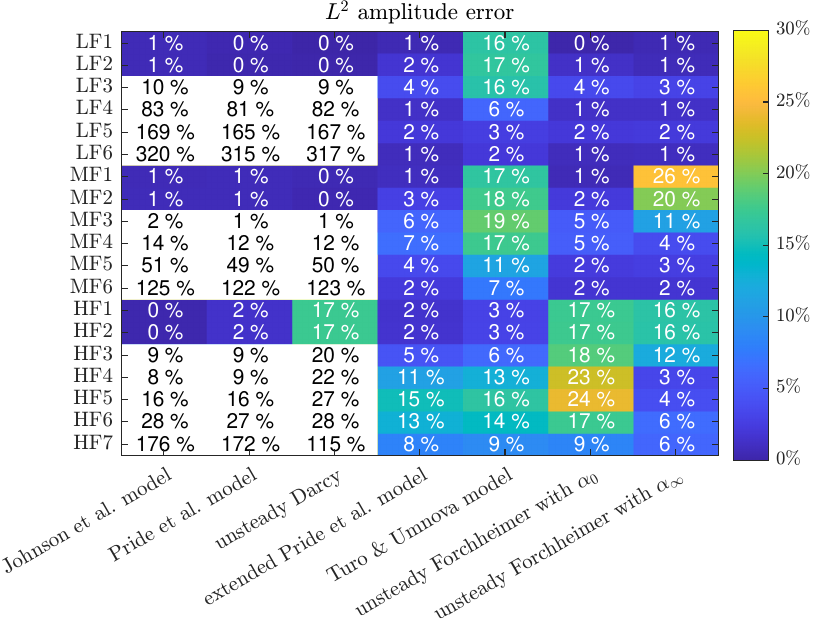}
    }%
    
    \subcaptionbox{}{
        \includegraphics[width=0.9\textwidth]{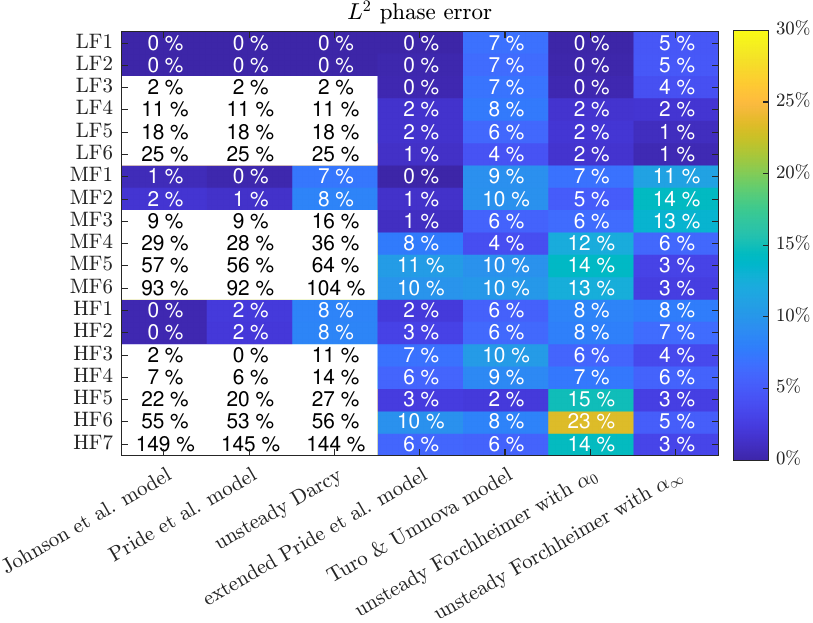}
    }
    
	\caption{Amplitude and phase contribution to the $L^2$ error normalised with $\max\superficialavg{u}$ of the respective flow case. The entries where a linear model would be applied to a nonlinear flow case are marked with a white background.}
	\label{tab:model_error_amplitude_phase}
\end{table}

\begin{figure}
	\centering
    \subcaptionbox{}{
        \includegraphics[width=0.45\textwidth]{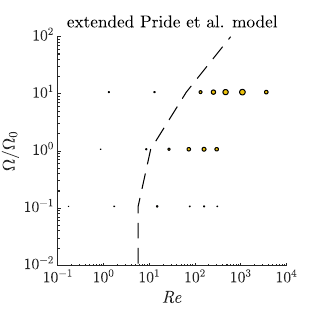}
    }
    \subcaptionbox{}{
        \includegraphics[width=0.45\textwidth]{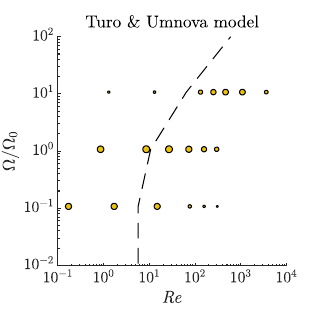}
    }%
    
    \subcaptionbox{}{
        \includegraphics[width=0.45\textwidth]{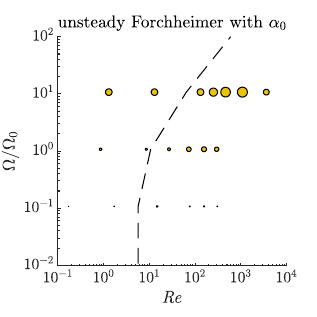}
    }
    \subcaptionbox{}{
        \includegraphics[width=0.45\textwidth]{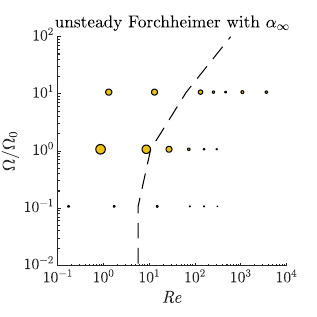}
    }
	\caption{Distribution of the $L^2$ model error in the $\Rey$--$\Omega/\Omega_0$ parameter space. The diameter of the circles is proportional to the $L^2$ error. The dashed line indicates the approximate boundary between linear and nonlinear flow \citep{Unglehrt.2022}. }
	\label{fig:model_error_parameter_space}
\end{figure}

The error of the model predictions with respect to the direct numerical simulation results is quantified using the cycle-averaged $L^2$ error \eqref{eq:L2_distance}; we further decompose this error into an amplitude \eqref{eq:amplitude_distance} and a phase contribution \eqref{eq:phase_distance} using a Fourier series. Table \ref{tab:model_error_amplitude_phase} shows the amplitude and phase contribution to the $L^2$ error of the different model predictions with respect to the direct numerical simulation dataset. Both the amplitude error and the phase error make significant contributions to the overall error of the models. The two components are weakly correlated, with the phase error being on average about $57 \%$ of the amplitude error, but the importance of the two components varies considerably between models and simulation cases.

For linear flow it can be seen that the dynamic permeability models of \citet{Johnson.1987} and \citet{Pride.1993} are very accurate over the entire frequency range whereas the unsteady Darcy equation of \citet{Zhu.2014} has very small errors at low frequencies and high errors at high frequencies. Thus, the behaviour of the linear models is consistent with the discussion in section \ref{sec:discussion_of_models}.

The prediction accuracy of the four nonlinear models shows significant differences depending on the flow case. This is illustrated in figure \ref{fig:model_error_parameter_space} which shows the variation of the $L^2$ error of the different nonlinear models over the $\Rey$--$\Wo$ parameter space.

We find that the extended dynamic permeability model based on the model of \citet{Pride.1993} has very small errors in linear flow and for flow at low frequencies. The reason for this is that the model equation \eqref{eq:extended_Pride_et_al_model} reverts to the model of \citet{Pride.1993} as $\Rey\to 0$ and to the steady Forchheimer equation as $\Omega/\Omega_0\to 0$. The prediction errors are largest for nonlinear flow at high frequencies. Interestingly, the prediction error for the case HF7 ($\Rey=3580$) is smaller than for the case HF6 ($\Rey=1080$). The model of \citet{Turo.2013} performs similar to the extended \citep{Pride.1993} model at high frequencies; however, the prediction errors in the medium and low frequency regime are very large. The reason for this is the excessive damping in the linear regime that was discussed in section \ref{sec:discussion_of_models}. The unsteady Forchheimer equation with the acceleration coefficient based on the static viscous tortuosity $\statictortuosity$ has relatively small errors at low frequencies, moderate errors at medium frequencies and very large errors at high frequencies. The largest errors can be observed in nonlinear flow at high frequencies.  In contrast, the unsteady Forchheimer equation with the acceleration coefficient based on the high-frequency limit of the dynamic tortuosity $\tortuosity$ has comparably small errors at low frequencies, very large errors in linear flow at medium and high frequencies and small errors for flow at high Reynolds numbers. This somewhat surprising behaviour of the unsteady Forchheimer equation with $\tortuosity$ will be investigated in the following section.

\subsection{Analysis of prediction errors}

\begin{figure}
    \centering
    \includegraphics[width=\textwidth]{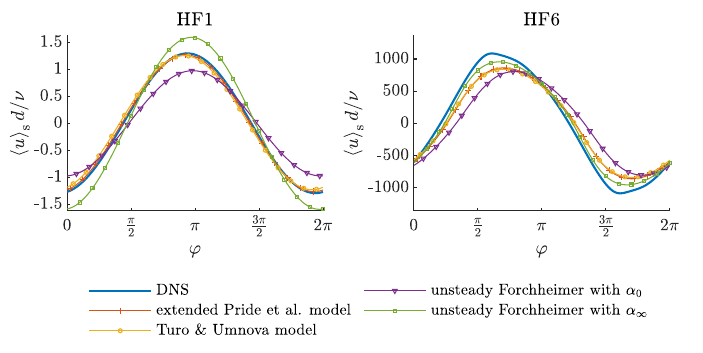}
    \caption{Comparison of model predictions of the superficial velocity for the cases HF1 and HF6.}
    \label{fig:comparison}
\end{figure}

In this section, we aim to explain the different error behaviours described above. We first investigate the discrepancies in the linear regime, where the model of \citet{Turo.2013} (equation \ref{eq:Turo_model}) and the unsteady Forchheimer equation \eqref{eq:unsteady_Darcy-Forchheimer_equation} have large errors at low and high frequencies, respectively. We then proceed to the nonlinear high frequency regime where large discrepancies between the models can be observed and the flow state is characterised by the interaction between strong accelerations and turbulence. Therefore, we take a closer look at the predictions of the models for the linear case HF1 and for the turbulent case HF6 at $\Omega/\Omega_0 = 10.7$ ($\Wo=100$). Figure \ref{fig:comparison} shows the predictions of the different models in comparison to the superficial velocity from the simulations.

For the linear case, the model of \citet{Turo.2013} (equation \ref{eq:Turo_model}) and the extended \citet{Pride.1993} model (equation \ref{eq:extended_Pride_et_al_model}) accurately represent the amplitude and phase of the superficial velocity; the errors in the maximum superficial velocity are $-3.1 \%$ and $-2.2 \%$, respectively. This is in agreement with the discussion in section \ref{sec:relations_among_the_linear_models} (figure \ref{fig:dyn_permeability}).  On the other hand, the unsteady Forchheimer equation \eqref{eq:unsteady_Darcy-Forchheimer_equation} with an acceleration coefficient $\statictortuosity$ underpredicts the amplitude of the superficial velocity by $24.8 \%$ whereas the unsteady Forchheimer equation with an acceleration coefficient $\tortuosity$ overpredicts the amplitude of the superficial velocity by $22.7 \%$.

Since for this case the magnitude of the nonlinear term is only about $1.5\%$ of the linear drag, the unsteady Forchheimer equation is effectively reduced to the unsteady Darcy equation here. Therefore, the behaviour is determined by the acceleration coefficient $c$ and the permeability (contained in the coefficient $a=\mu/\permeability)$). The comparison of the two different choices for the acceleration coefficient $c$ of the unsteady Forchheimer equation thus highlights the effect of the ratio between inertia and linear drag on the superficial velocity. While the choice $c=\rho\,\tortuosity/\porosity$ appears to have too little mass, the choice $c=\rho\,\statictortuosity/\porosity$ appears to have too much mass. Since the behaviour of the case HF1 is sinusoidal, it would be possible to find coefficients for the unsteady Darcy equation that represent the simulation data exactly. However, these best fit coefficients would only be valid at this particular frequency and lead to inconsistent behaviour in the low- and high frequency limit (see also figure \ref{fig:dyn_permeability}). Choosing the coefficients as a function of the frequency leads to the frequency domain formulation of the dynamic permeability models (see section \ref{sec:dynamic_permeability_models}). In the time domain, this is reflected in the appearance of a history term.

For the turbulent case, the amplitude of the superficial velocity in the direct numerical simulation is $16 \%$ lower than the amplitude of the linear case HF1 scaled by a factor of $10^3$ to the same Hagen number. Moreover, the maximum superficial velocity in the case HF6 is attained significantly earlier than in the case HF1 ($\varphi=0.61\pi$ for HF6 compared to $\varphi=0.94\pi$ for HF1). It can be seen in figure \ref{fig:comparison} that all models underpredict the amplitude of the superficial velocity: The error in the maximum superficial velocity is $-23\%$ for the model of \citet{Turo.2013}, $-21.2\%$ for the extended \citet{Pride.1993} model, $-26.4\%$ for the unsteady Forchheimer equation with $\statictortuosity$ and $-12.9\%$ for the unsteady Forchheimer equation with $\tortuosity$. The phase is captured satisfactorily by the extended \citet{Pride.1993} model, the model of \citet{Turo.2013} and the unsteady Forchheimer equation with $\tortuosity$, whereas the unsteady Forchheimer equation with $\statictortuosity$ mispredicts the phase. This is reflected clearly in the phase error (table \ref{tab:model_error_amplitude_phase}). Furthermore, figure \ref{fig:comparison} shows that all models fail to represent the relatively sharp bend at the peak.

\begin{figure}
    \centering
    \includegraphics[width=\textwidth]{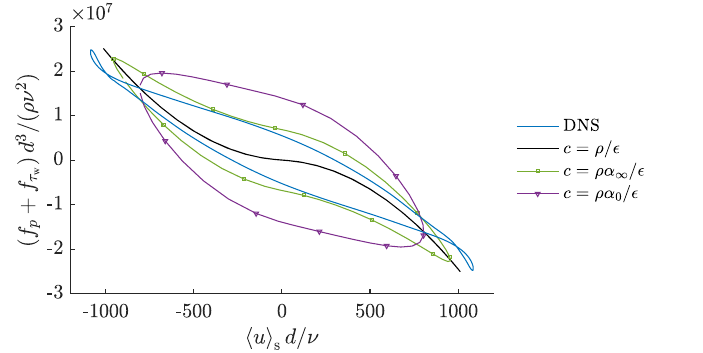}
    \caption{Comparison of the drag from the direct numerical simulation of the case HF6 with the drag from model solutions to the unsteady Forchheimer equation for the values of the acceleration coefficient $c=\rho/\porosity$ (quasi-steady closure / zero virtual mass), $c=\rho\,\tortuosity/\porosity$ (high-frequency limit of the dynamic tortuosity) and $c=\rho\,\statictortuosity/\porosity$ (static viscous tortuosity).}
    \label{fig:nonlinear_drag}
\end{figure}

It can be seen from the comparison to the linear case that the (identical) nonlinear term in the models causes excessive damping, since even the unsteady Forchheimer equation with an acceleration coefficient $c=\rho\,\tortuosity/\porosity$ underpredicts the amplitude of the simulation. In this sense, the behaviour of the different models could be understood as a compensation or superposition of errors.

This interpretation is supported by figure \ref{fig:nonlinear_drag}, which compares the drag of the direct numerical simulation of the case HF6 to the drag of model solutions by the unsteady Forchheimer equation for different values of the acceleration coefficient. The drag is determined for each time series according to the volume-averaged momentum equation \eqref{eq:avg_momentum_equation} as the sum of the acceleration and the pressure gradient. We first consider the curve for the value $c=\rho/\porosity$, which corresponds to the case of zero virtual mass and was assumed for instance by \citet{Sollitt.1972}, \citet{Kuznetsov.2006} or \citet{Breugem.2006}. It can be seen that the drag is a single-valued function of the superficial velocity
\begin{equation}
    f_p+f_{\tau_{\mathrm{w}x}} = -\porosity \left(a \superficialavg{u} + b\left\vert\superficialavg{u}\right\vert\superficialavg{u}\right) \,.
\end{equation}
that is identical to the drag in the steady flow \eqref{eq:Forchheimer_equation}. For higher values of the acceleration coefficient, the drag becomes a double-valued function of the superficial velocity. Since the drag in the direct numerical simulation is also multi-valued, a nonzero virtual mass ($c>\rho/\porosity$) is required to represent the correct behaviour. It can be seen that the peak superficial velocity of the model predictions decreases as the acceleration coefficient increases. The peak superficial velocity of the simulation cannot be reached even for $c=\rho/\porosity$ for which the highest amplitude of the different model predictions is obtained. As $b\left\vert \superficialavg{u}\right\vert \approx 12 a$, this indicates that the nonlinear drag is too large.

It is important to realise that the physical assumptions underlying the different models are not fulfilled in this flow case, for instance:
\begin{itemize}
    \item The unsteady Forchheimer equation implies an energy equation (see section \ref{sec:implied_energy_equation}) in which both the kinetic energy and the dissipation rate are single-valued functions of the instantaneous superficial velocity. This is not the case in the direct numerical simulation: The kinetic energy during acceleration is significantly lower than during deceleration due to the generation of turbulent kinetic energy and the kinetic energy is not in phase with the dissipation rate  \citep{Unglehrt.2023_DLES} -- as it would have to be if both were single-valued functions of the superficial velocity.
    \item The model of \citet{Turo.2013} (and similarly the model of \citet{Pride.1993}) assume linear Stokes boundary layers at high frequencies that evolve according to an outer potential flow unlimited by convection. However, when the flow separates, the outer flow is modified such that the boundary layers would have to evolve differently, resulting in a different formulation of the history term.
\end{itemize}
Therefore, different parametrisations of the drag should be explored beyond the models investigated here.

In conclusion, we could explain the mispredictions of the unsteady Forchheimer equation for linear flow at high frequencies. These could be attributed mainly to a mismatch of the ratio between the inertia and the linear drag. A consistent resolution of this issue is provided by the dynamic permeability models that choose this ratio depending on the frequency. The behaviour of the models in nonlinear flow at high frequencies could be partially explained by an overprediction of the nonlinear drag. However, since many intrinsic assumptions of the models are violated in this regime, the overall functional form of the parametrisations should be revisited.

\section{Conclusion}

In this contribution, we reviewed various models for unsteady porous media flow from the literature and compared their predictions for oscillatory flow through a hexagonal sphere pack. The reference data are direct numerical simulations of \citet{Zhu.2016a,Unglehrt.2022,Unglehrt.2023,Unglehrt.2023_DLES} for this flow configuration. 

The models can be divided into two classes: On the one hand, there are the unsteady variants of the Darcy and Forchheimer equation \citep{Polubarinova-Kochina.1962}; on the other hand, there are the dynamic permeability models \citep{Johnson.1987,Pride.1993} which feature a convolution-type structure in the time domain and nonlinear extensions thereof \citep{Turo.2013}.

In linear flow, the dynamic permeability models of \citet{Johnson.1987,Pride.1993} provide an accurate description of the simulation data. The unsteady Darcy equation could be obtained as a special case of these models in which the acceleration coefficient is either based on the static viscous tortuosity $\statictortuosity$ or on the high-frequency limit of the dynamic tortuosity $\tortuosity$. The model of \citet{Turo.2013} is overly dissipative at medium and low frequencies. To alleviate this drawback, we constructed a similar model based on the model of \citet{Pride.1993}.

In nonlinear flow, we compared the unsteady Forchheimer equation for two different choices of the acceleration coefficient, the extended dynamic permeability model of \citet{Turo.2013} and an analogous formulation based on the model of \citet{Pride.1993}. The unsteady Forchheimer equation with the acceleration coefficient based on the static viscous tortuosity $\statictortuosity$ shows good results in the low frequency regime, but the results deteriorate at higher frequencies and are particularly bad in the high frequency nonlinear regime. On the other hand, the unsteady Forchheimer equation with the acceleration coefficient chosen based on the high-frequency limit of the dynamic tortuosity $\tortuosity$ shows very good results at high Reynolds numbers, but incurs large errors in the linear regime. The model of \citet{Turo.2013} has relatively large errors throughout the entire parameter space. On the other hand, the extension of the model of \citet{Pride.1993} shows excellent results in the linear regime and at low frequencies, but the results deteriorate for nonlinear flow at high frequencies.

Generally, our proposed extension of the model of \citet{Pride.1993} along the lines of \citep{Turo.2013} (see section \ref{sec:improvement}) seems to be a robust choice with accurate results in linear unsteady and nonlinear steady flow and a moderate increase of errors towards strongly accelerated nonlinear flow. The drawback of this model is the additional implementation effort and computational cost caused by the convolution term. For weakly accelerated flow, this model can be simplified to the unsteady Forchheimer equation with an acceleration coefficient $c=\rho\,\statictortuosity/\porosity$ based on the static viscous tortuosity $\statictortuosity$, which is more economical. On the other hand, the unsteady Forchheimer equation with an acceleration coefficient $c=\rho\,\tortuosity/\porosity$ based on the high frequency limit of the dynamic tortuosity $\tortuosity$ should be used judiciously since the small errors at large Reynolds numbers must be weighed against large errors for linear unsteady flow.

Further improvements are needed in the parametrisation of the nonlinear drag at high frequencies, as our results indicate that the Forchheimer term leads to an overprediction of the nonlinear drag. Moreover, our previous investigations showed that there is a phase lag between the nonlinear effects in the velocity field and in the drag and the superficial velocity \citep{Unglehrt.2022,Unglehrt.2023}. It thus seems plausible that introducing a time lag between the nonlinear drag and the superficial velocity could lead to an improved model. Further research should also aim to generalise the present results to different kinds of porous media, for example random and polydisperse sphere packs, foams, and cylinder arrays.

\backmatter

\begin{appendices}

\end{appendices}

\section*{Declarations}

\bmhead{Funding}
The authors gratefully acknowledge the financial support of the Deutsche Forschungsgemeinschaft
(DFG, German Research Foundation) under grant no. MA2062/13-1. Computing time was granted by the
Leibniz Supercomputing Centre on its Linux-Cluster.

\bmhead{Competing Interests}
The authors have no relevant financial or non-financial interests to disclose.

\bmhead{Author Contributions}
L.U. performed the review of the models from the literature, implemented the models and analysed the simulation data. L.U. and M.M. both contributed to writing the manuscript.

\bmhead{Data Availability}
The time series of the superficial velocity and kinetic energy for the simulations LF1--LF4, MF1--MF4 and HF1--HF4 are provided as a supplement to \citep{Unglehrt.2022}. The time series of the superficial velocity and kinetic energy for the simulations LF5, LF6, MF5, MF6 and HF5 are provided as a supplement to \citep{Unglehrt.2023}. The time series of the superficial velocity and kinetic energy for the simulations HF6 and HF7  are provided as a supplement to this work.

\bibliography{references}

\begin{thebibliography}{76}
\providecommand{\natexlab}[1]{#1}
\providecommand{\url}[1]{{#1}}
\providecommand{\urlprefix}{URL }
\providecommand{\doi}[1]{\url{https://doi.org/#1}}
\providecommand{\eprint}[2][]{\url{#2}}
 \bibcommenthead

\bibitem[{Aboujafari et~al(2022)Aboujafari, Valipour, Hajialimohammadi, and
  Honnery}]{Aboujafari.2022}
Aboujafari M, Valipour MS, Hajialimohammadi A, et~al (2022) Porous {{Medium
  Applications}} in {{Internal Combustion Engines}}: {{A Review}}. Transport in
  Porous Media 141(3):799--824. \doi{10.1007/s11242-022-01750-2}

\bibitem[{Andreades et~al(2014)Andreades, Cisneros, Choi, Chong, Fratoni, Hong,
  Huddar, Huff, Krumwiede, Laufer, Munk, Scarlat, Zweibaum, Greenspan, and
  Peterson}]{Andreades.2014}
Andreades CH, Cisneros AT, Choi JK, et~al (2014) Technical {{Description}} of
  the ``{{Mark}} 1'' {{Pebble-Bed Fluoride-Salt-Cooled High-Temperature
  Reactor}} ({{PB-FHR}}) {{Power Plant}}. Tech. Rep. UCBTH- 14- 002,
  {Department of Nuclear Engineering University of California, Berkeley},
  {Berkeley}

\bibitem[{Batchelor(2000)}]{Batchelor.2000}
Batchelor GK (2000) An {{Introduction}} to {{Fluid Dynamics}}. {Cambridge
  University Press}, {Cambridge}, \doi{10.1017/CBO9780511800955}

\bibitem[{Boutin and Geindreau(2010)}]{Boutin.2010}
Boutin C, Geindreau C (2010) Periodic homogenization and consistent estimates
  of transport parameters through sphere and polyhedron packings in the whole
  porosity range. Physical Review E 82(3):036,313.
  \doi{10.1103/PhysRevE.82.036313}

\bibitem[{Breugem et~al(2006)Breugem, Boersma, and
  Uittenbogaard}]{Breugem.2006}
Breugem WP, Boersma BJ, Uittenbogaard RE (2006) The influence of wall
  permeability on turbulent channel flow. Journal of Fluid Mechanics 562:35.
  \doi{10.1017/S0022112006000887}

\bibitem[{Burcharth and Andersen(1995)}]{Burcharth.1995}
Burcharth H, Andersen O (1995) On the one-dimensional steady and unsteady
  porous flow equations. Coastal Engineering 24(3-4):233--257.
  \doi{10.1016/0378-3839(94)00025-S}

\bibitem[{Champoux and Allard(1991)}]{Champoux.1991}
Champoux Y, Allard JF (1991) Dynamic tortuosity and bulk modulus in
  air-saturated porous media. Journal of Applied Physics 70(4):1975--1979.
  \doi{10.1063/1.349482}

\bibitem[{Chapman and Higdon(1992)}]{Chapman.1992}
Chapman AM, Higdon JJL (1992) Oscillatory {{Stokes}} flow in periodic porous
  media. Physics of Fluids A: Fluid Dynamics 4(10):2099--2116.
  \doi{10.1063/1.858507}

\bibitem[{Chorin(1968)}]{Chorin.1968}
Chorin AJ (1968) Numerical solution of the {{Navier-Stokes}} equations.
  Mathematics of Computation 22(104):745--762.
  \doi{10.1090/S0025-5718-1968-0242392-2}

\bibitem[{Cortis et~al(2002)Cortis, Smeulders, Lafarge, Firdaouss, and
  Guermond}]{Cortis.2002}
Cortis A, Smeulders DMJ, Lafarge D, et~al (2002) Geometry {{Effects}} on
  {{Sound}} in {{Porous Media}}. In: Ehlers W (ed) {{IUTAM Symposium}} on
  {{Theoretical}} and {{Numerical Methods}} in {{Continuum Mechanics}} of
  {{Porous Materials}}, vol~87. {Kluwer Academic Publishers}, {Dordrecht}, p
  187--192, \doi{10.1007/0-306-46953-7_26}

\bibitem[{Cortis et~al(2003)Cortis, Smeulders, Guermond, and
  Lafarge}]{Cortis.2003}
Cortis A, Smeulders DMJ, Guermond JL, et~al (2003) Influence of pore roughness
  on high-frequency permeability. Physics of Fluids 15(6):1766--1775.
  \doi{10.1063/1.1571545}

\bibitem[{Darcy(1856)}]{Darcy.1856}
Darcy H (1856) {Les fontaines publiques de la ville de Dijon}. {Victor
  Dalmont}, {Paris}

\bibitem[{Davit et~al(2013)Davit, Bell, Byrne, Chapman, Kimpton, Lang, Leonard,
  Oliver, Pearson, Shipley, Waters, Whiteley, Wood, and Quintard}]{Davit.2013}
Davit Y, Bell CG, Byrne HM, et~al (2013) Homogenization via formal multiscale
  asymptotics and volume averaging: {{How}} do the two techniques compare?
  Advances in Water Resources 62:178--206.
  \doi{10.1016/j.advwatres.2013.09.006}

\bibitem[{Diethelm et~al(2005)Diethelm, Ford, Freed, and
  Luchko}]{Diethelm.2005}
Diethelm K, Ford N, Freed A, et~al (2005) Algorithms for the fractional
  calculus: {{A}} selection of numerical methods. Computer Methods in Applied
  Mechanics and Engineering 194(6-8):743--773. \doi{10.1016/j.cma.2004.06.006}

\bibitem[{Ene and {Sanchez-Palencia}(1975)}]{Ene.1975}
Ene H, {Sanchez-Palencia} E (1975) Equations et ph\'enom\`enes de surface pour
  l'\'ecoulement dans un milieu poreux. Journal de M\'ecanique 14

\bibitem[{Ergun(1952)}]{Ergun.1952}
Ergun S (1952) Fluid {{Flow Through Packed Columns}}. Chemical Engineering
  Progress 48(2):89--94

\bibitem[{Firdaouss et~al(1997)Firdaouss, Guermond, and
  Le~Qu{\'e}r{\'e}}]{Firdaouss.1997}
Firdaouss M, Guermond JL, Le~Qu{\'e}r{\'e} P (1997) Nonlinear corrections to
  {{Darcy}}'s law at low {{Reynolds}} numbers. Journal of Fluid Mechanics
  343:331--350. \doi{10.1017/S0022112097005843}

\bibitem[{Forchheimer(1901)}]{Forchheimer.1901}
Forchheimer P (1901) Wasserbewegung durch {{Boden}}. Zeitschrift des Vereins
  deutscher Ingenieure 45:1782--1788

\bibitem[{Graham and Higdon(2002)}]{Graham.2002}
Graham DR, Higdon JJL (2002) Oscillatory forcing of flow through porous media.
  {{Part}} 2. {{Unsteady}} flow. Journal of Fluid Mechanics 465.
  \doi{10.1017/s0022112002001143}

\bibitem[{Gu and Wang(1991)}]{Gu.1991}
Gu Z, Wang H (1991) Gravity waves over porous bottoms. Coastal Engineering
  15(5-6):497--524. \doi{10.1016/0378-3839(91)90025-C}

\bibitem[{Hall et~al(1995)Hall, Smith, and Turcke}]{Hall.1995}
Hall KR, Smith GM, Turcke DJ (1995) Comparison of oscillatory and stationary
  flow through porous media. Coastal Engineering 24(3-4):217--232.
  \doi{10.1016/0378-3839(94)00017-R}

\bibitem[{Hill et~al(2001)Hill, Koch, and Ladd}]{Hill.2001}
Hill RJ, Koch DL, Ladd AJC (2001) The first effects of fluid inertia on flows
  in ordered and random arrays of spheres. Journal of Fluid Mechanics
  448:213--241. \doi{10.1017/S0022112001005948}

\bibitem[{Hsu and Cheng(1990)}]{Hsu.1990}
Hsu C, Cheng P (1990) Thermal dispersion in a porous medium. International
  Journal of Heat and Mass Transfer 33(8):1587--1597.
  \doi{10.1016/0017-9310(90)90015-M}

\bibitem[{Iliuta and Larachi(2016)}]{Iliuta.2016}
Iliuta I, Larachi F (2016) Three-dimensional simulations of gas-liquid
  cocurrent downflow in vertical, inclined, and oscillating packed beds. AIChE
  Journal 62(3):916--927. \doi{10.1002/aic.15071}

\bibitem[{Iliuta and Larachi(2017)}]{Iliuta.2017}
Iliuta I, Larachi F (2017) {{CO2}} abatement in oscillating packed-bed
  scrubbers: {{Hydrodynamics}} and reaction performances for marine
  applications. AIChE Journal 63(3):1064--1076. \doi{10.1002/aic.15450}

\bibitem[{Johnson et~al(1987)Johnson, Koplik, and Dashen}]{Johnson.1987}
Johnson DL, Koplik J, Dashen R (1987) Theory of dynamic permeability and
  tortuosity in fluid-saturated porous media. Journal of Fluid Mechanics
  176:379--402. \doi{10.1017/S0022112087000727}

\bibitem[{Kahler and Kabala(2019)}]{Kahler.2019}
Kahler DM, Kabala ZJ (2019) Acceleration of {{Groundwater Remediation}} by
  {{Rapidly Pulsed Pumping}}: {{Laboratory Column Tests}}. Journal of
  Environmental Engineering 145(1):06018,009.
  \doi{10.1061/(ASCE)EE.1943-7870.0001479}

\bibitem[{Kergomard et~al(2013)Kergomard, Lafarge, and
  Gilbert}]{Kergomard.2013}
Kergomard J, Lafarge D, Gilbert J (2013) Transients in {{Porous Media}}:
  {{Exact}} and {{Modelled Time-Domain Green}}'s {{Functions}}. Acta Acustica
  united with Acustica 99(4):557--571. \doi{10.3813/AAA.918635}

\bibitem[{Koch and Ladd(1997)}]{Koch.1997}
Koch DL, Ladd AJC (1997) Moderate {{Reynolds}} number flows through periodic
  and random arrays of aligned cylinders. Journal of Fluid Mechanics
  349:31--66. \doi{10.1017/S002211209700671X}

\bibitem[{Kuznetsov and Nield(2006)}]{Kuznetsov.2006}
Kuznetsov AV, Nield DA (2006) Forced {{Convection}} with {{Laminar Pulsating
  Flow}} in a {{Saturated Porous Channel}} or {{Tube}}. Transport in Porous
  Media 65(3):505--523. \doi{10.1007/s11242-006-6791-6}

\bibitem[{Lafarge(1993)}]{Lafarge.1993}
Lafarge D (1993) {Propagation du son dans les mat\'eriaux poreux \'a structure
  rig ide satur\'es par un fluide viscothermique: D\'efinition de param\`etres
  g\'eom\'etriques, analogie electromagn\'etique, temps de relaxation}. PhD
  thesis, Universit\'e du Maine, {Le Mans}

\bibitem[{Lafarge(2009)}]{Lafarge.2009}
Lafarge D (2009) The {{Equivalent Fluid Model}}. In: Bruneau M, Potel C (eds)
  Materials and {{Acoustics Handbook}}. {ISTE}, {London, UK}, p 153--204,
  \doi{10.1002/9780470611609.ch6}

\bibitem[{Lasseux et~al(2011)Lasseux, Abbasian~Arani, and
  Ahmadi}]{Lasseux.2011}
Lasseux D, Abbasian~Arani AA, Ahmadi A (2011) On the stationary macroscopic
  inertial effects for one phase flow in ordered and disordered porous media.
  Physics of Fluids 23(7):073,103. \doi{10.1063/1.3615514}

\bibitem[{Losada et~al(1995)Losada, Losada, and Mart{\'i}n}]{Losada.1995}
Losada I, Losada M, Mart{\'i}n F (1995) Experimental study of wave-induced flow
  in a porous structure. Coastal Engineering 26(1-2):77--98.
  \doi{10.1016/0378-3839(95)00013-5}

\bibitem[{Lowe et~al(2008)Lowe, Shavit, Falter, Koseff, and
  Monismith}]{Lowe.2008}
Lowe RJ, Shavit U, Falter JL, et~al (2008) Modeling flow in coral communities
  with and without waves: {{A}} synthesis of porous media and canopy flow
  approaches. Limnology and Oceanography 53(6):2668--2680.
  \doi{10.4319/lo.2008.53.6.2668}

\bibitem[{Macdonald et~al(1979)Macdonald, {El-Sayed}, Mow, and
  Dullien}]{Macdonald.1979}
Macdonald IF, {El-Sayed} MS, Mow K, et~al (1979) Flow through {{Porous
  Media-the Ergun Equation Revisited}}. Industrial \& Engineering Chemistry
  Fundamentals 18(3):199--208. \doi{10.1021/i160071a001}

\bibitem[{Manhart et~al(2001)Manhart, Tremblay, and Friedrich}]{Manhart.2001}
Manhart M, Tremblay F, Friedrich R (2001) {{MGLET}}: A parallel code for
  efficient {{DNS}} and {{LES}} of complex geometries. In: Parallel
  {{Computational Fluid Dynamics}} 2000. {North-Holland}, {Amsterdam}, p
  449--456, \doi{10.1016/B978-044450673-3/50123-8}

\bibitem[{Mei and Auriault(1991)}]{Mei.1991}
Mei CC, Auriault JL (1991) The effect of weak inertia on flow through a porous
  medium. Journal of Fluid Mechanics 222:647--663.
  \doi{10.1017/S0022112091001258}

\bibitem[{Muttray(2000)}]{Muttray.2000}
Muttray M (2000) {Wellenbewegung an und in einem gesch\"utteten Wellenbrecher -
  Laborexperimente im Gro\ss ma\ss stab und theoretische Untersuchungen}. PhD
  thesis, Technische Universit\"at Braunschweig, {Braunschweig}

\bibitem[{Ni et~al(2003)Ni, Mackley, Harvey, Stonestreet, Baird, and
  Rama~Rao}]{Ni.2003}
Ni X, Mackley M, Harvey A, et~al (2003) Mixing {{Through Oscillations}} and
  {{Pulsations}}\textemdash{{A Guide}} to {{Achieving Process Enhancements}} in
  the {{Chemical}} and {{Process Industries}}. Chemical Engineering Research
  and Design 81(3):373--383. \doi{10.1205/02638760360596928}

\bibitem[{Nield(1991)}]{Nield.1991}
Nield D (1991) The limitations of the {{Brinkman-Forchheimer}} equation in
  modeling flow in a saturated porous medium and at an interface. International
  Journal of Heat and Fluid Flow 12(3):269--272.
  \doi{10.1016/0142-727X(91)90062-Z}

\bibitem[{Nield(2000)}]{Nield.2000}
Nield DA (2000) Resolution of a {{Paradox Involving Viscous Dissipation}} and
  {{Nonlinear Drag}} in a {{Porous Medium}}. Transport in Porous Media
  41(3):349--357. \doi{10.1023/A:1006636605498}

\bibitem[{Norris(1986)}]{Norris.1986}
Norris AN (1986) On the viscodynamic operator in {{Biot}}'s equations of
  poroelasticity. Journal of Wave-material Interaction 1:365--380

\bibitem[{{Pa{\'e}z-Garc{\'i}a} et~al(2017){Pa{\'e}z-Garc{\'i}a},
  {Vald{\'e}s-Parada}, and Lasseux}]{Paez-Garcia.2017}
{Pa{\'e}z-Garc{\'i}a} CT, {Vald{\'e}s-Parada} FJ, Lasseux D (2017) Macroscopic
  momentum and mechanical energy equations for incompressible single-phase flow
  in porous media. Physical Review E 95(2). \doi{10.1103/PhysRevE.95.023101}

\bibitem[{Peller(2010)}]{Peller.2010}
Peller N (2010) Numerische {{Simulation}} turbulenter {{Str\"omungen}} mit
  {{Immersed Boundaries}}. PhD thesis, Technische Universit\"at M\"unchen,
  {M\"unchen}

\bibitem[{Peller et~al(2006)Peller, Duc, Tremblay, and Manhart}]{Peller.2006}
Peller N, Duc AL, Tremblay F, et~al (2006) High-order stable interpolations for
  immersed boundary methods. International Journal for Numerical Methods in
  Fluids 52(11):1175--1193. \doi{10.1002/fld.1227}

\bibitem[{{Polubarinova-Kochina}(1962)}]{Polubarinova-Kochina.1962}
{Polubarinova-Kochina} PI (1962) Theory of {{Ground Water Movement}}.
  {Princeton University Press}, {Princeton, New Jersey}

\bibitem[{Pride et~al(1993)Pride, Morgan, and Gangi}]{Pride.1993}
Pride SR, Morgan FD, Gangi AF (1993) Drag forces of porous-medium acoustics.
  Physical Review B 47(9):4964--4978. \doi{10.1103/PhysRevB.47.4964}

\bibitem[{Roncen et~al(2018)Roncen, Fellah, Lafarge, Piot, Simon, Ogam, Fellah,
  and Depollier}]{Roncen.2018}
Roncen R, Fellah ZEA, Lafarge D, et~al (2018) Acoustical modeling and
  {{Bayesian}} inference for rigid porous media in the low-mid frequency
  regime. The Journal of the Acoustical Society of America 144(6):3084--3101.
  \doi{10.1121/1.5080561}

\bibitem[{Sakai and Manhart(2020)}]{Sakai.2020}
Sakai Y, Manhart M (2020) Consistent {{Flow Structure Evolution}} in
  {{Accelerating Flow Through Hexagonal Sphere Pack}}. Flow, Turbulence and
  Combustion 105(2):581--606. \doi{10.1007/s10494-020-00168-4}

\bibitem[{Schumann(1977)}]{Schumann.1977}
Schumann U (1977) Realizability of {{Reynolds-Stress Turbulence Models}}.
  Physics of Fluids 20:721--725. \doi{10.1063/1.861942}

\bibitem[{Shampine and Reichelt(1997)}]{Shampine.1997}
Shampine L, Reichelt M (1997) The {{MATLAB ODE Suite}}. SIAM J Sci Comput
  \doi{10.1137/S1064827594276424}

\bibitem[{Simon and Seume(1988)}]{Simon.1988}
Simon TW, Seume JR (1988) A {{Survey}} of {{Oscillating Engine Heat
  Exchangers}}. Tech. Rep. NASA-CR-182108), {University of Minnesota},
  {Minneapolis, Minnesota}

\bibitem[{Smeulders et~al(1992)Smeulders, Eggels, and
  Van~Dongen}]{Smeulders.1992}
Smeulders DMJ, Eggels RLGM, Van~Dongen MEH (1992) Dynamic permeability:
  Reformulation of theory and new experimental and numerical data. Journal of
  Fluid Mechanics 245(-1):211. \doi{10.1017/S0022112092000429}

\bibitem[{Sollitt and Cross(1972)}]{Sollitt.1972}
Sollitt CK, Cross RH (1972) Wave {{Transmission}} through {{Permeable
  Breakwaters}}. In: Coastal {{Engineering}} 1972. {American Society of Civil
  Engineers}, {Vancouver, British Columbia, Canada}, pp 1827--1846,
  \doi{10.1061/9780872620490.106}

\bibitem[{Trevizoli et~al(2016)Trevizoli, Peixer, and Barbosa}]{Trevizoli.2016}
Trevizoli PV, Peixer GF, Barbosa JR (2016) Thermal\textendash hydraulic
  evaluation of oscillating-flow regenerators using water: {{Experimental}}
  analysis of packed beds of spheres. International Journal of Heat and Mass
  Transfer 99:918--930. \doi{10.1016/j.ijheatmasstransfer.2016.03.014}

\bibitem[{Turo and Umnova(2013)}]{Turo.2013}
Turo D, Umnova O (2013) Influence of {{Forchheimer}}'s nonlinearity and
  transient effects on pulse propagation in air saturated rigid granular
  materials. The Journal of the Acoustical Society of America
  134(6):4763--4774. \doi{10.1121/1.4824969}

\bibitem[{Umnova and Turo(2009)}]{Umnova.2009}
Umnova O, Turo D (2009) Time domain formulation of the equivalent fluid model
  for rigid porous media. The Journal of the Acoustical Society of America
  125(4):1860--1863. \doi{10.1121/1.3082123}

\bibitem[{Unglehrt(2024)}]{Unglehrt.2024}
Unglehrt L (2024) Oscillatory flow through porous media. PhD thesis, Technische
  Universit{\"a}t M{\"u}nchen,
  \urlprefix\url{https://nbn-resolving.de/urn/resolver.pl?urn:nbn:de:bvb:91-diss-20240422-1729913-1-2}

\bibitem[{Unglehrt and Manhart(2022{\natexlab{a}})}]{Unglehrt.2022}
Unglehrt L, Manhart M (2022{\natexlab{a}}) Onset of nonlinearity in oscillatory
  flow through a hexagonal sphere pack. Journal of Fluid Mechanics 944:A30.
  \doi{10.1017/jfm.2022.496}

\bibitem[{Unglehrt and Manhart(2022{\natexlab{b}})}]{Unglehrt.2022_TSFP}
Unglehrt L, Manhart M (2022{\natexlab{b}}) Symmetry {{Breaking}} and
  {{Turbulence}} in {{Oscillatory Flow Through}} a {{Hexagonal Sphere Pack}}.
  In: Proceedings of {{TSFP-12}} (2022) {{Osaka}}, {Osaka, Japan}, p~6

\bibitem[{Unglehrt and Manhart(2023{\natexlab{a}})}]{Unglehrt.2023}
Unglehrt L, Manhart M (2023{\natexlab{a}}) Decomposition of the drag force in
  steady and oscillatory flow through a hexagonal sphere pack. Journal of Fluid
  Mechanics 974:A32. \doi{10.1017/jfm.2023.798}

\bibitem[{Unglehrt and Manhart(2023{\natexlab{b}})}]{Unglehrt.2023_DLES}
Unglehrt L, Manhart M (2023{\natexlab{b}}) Direct and
  {{Large}}\textendash{{Eddy}} simulation of turbulent oscillatory flow through
  a hexagonal sphere pack. In: Marchioli C, Salvetti MV, {Garcia-Villalba} M,
  et~al (eds) Direct and {{Large Eddy Simulation XIII}}: {{Proceedings}} of
  {{DLES13}}, 1st edn. No.~31 in {{ERCOFTAC Series}}, {Springer Cham}, p
  118--123

\bibitem[{{van Gent}(1993)}]{vanGent.1993}
{van Gent} MRA (1993) Stationary and oscillatory flow through coarse porous
  media. Communications on hydraulic and geotechnical engineering, No 1993-09

\bibitem[{{van Gent}(1994)}]{vanGent.1994}
{van Gent} MRA (1994) The modelling of wave action on and in coastal
  structures. Coastal Engineering 22(3-4):311--339.
  \doi{10.1016/0378-3839(94)90041-8}

\bibitem[{{van Hinsberg} et~al(2011){van Hinsberg}, {ten Thije Boonkkamp}, and
  Clercx}]{vanHinsberg.2011}
{van Hinsberg} MAT, {ten Thije Boonkkamp} JHM, Clercx HJH (2011) An efficient,
  second order method for the approximation of the {{Basset}} history force.
  Journal of Computational Physics 230(4):1465--1478.
  \doi{10.1016/j.jcp.2010.11.014}

\bibitem[{Verstappen and Veldman(2003)}]{Verstappen.2003}
Verstappen R, Veldman A (2003) Symmetry-preserving discretization of turbulent
  flow. Journal of Computational Physics 187(1):343--368.
  \doi{10.1016/S0021-9991(03)00126-8}

\bibitem[{Wang(2008)}]{Wang.2008}
Wang CY (2008) The {{Starting Flow}} in {{Ducts Filled}} with a
  {{Darcy}}\textendash{{Brinkman Medium}}. Transport in Porous Media
  75(1):55--62. \doi{10.1007/s11242-008-9210-3}

\bibitem[{Ward(1964)}]{Ward.1964}
Ward JC (1964) Turbulent {{Flow}} in {{Porous Media}}. Journal of the
  Hydraulics Division 90(5):1--12. \doi{10.1061/JYCEAJ.0001096}

\bibitem[{Whitaker(1967)}]{Whitaker.1967}
Whitaker S (1967) Diffusion and dispersion in porous media. AIChE Journal
  13(3):420--427. \doi{10.1002/aic.690130308}

\bibitem[{Whitaker(1986)}]{Whitaker.1986}
Whitaker S (1986) Flow in porous media {{I}}: {{A}} theoretical derivation of
  {{Darcy}}'s law. Transport in Porous Media 1(1):3--25.
  \doi{10.1007/BF01036523}

\bibitem[{Whitaker(1996)}]{Whitaker.1996}
Whitaker S (1996) The {{Forchheimer}} equation: {{A}} theoretical development.
  Transport in Porous Media 25(1):27--61. \doi{10.1007/BF00141261}

\bibitem[{Williamson(1980)}]{Williamson.1980}
Williamson J (1980) Low-storage {{Runge-Kutta}} schemes. Journal of
  Computational Physics 35(1):48--56. \doi{10.1016/0021-9991(80)90033-9}

\bibitem[{Zhu(2016)}]{Zhu.2016}
Zhu T (2016) Unsteady porous-media flows. PhD thesis, Technische Universit\"at
  M\"unchen, {M\"unchen}

\bibitem[{Zhu and Manhart(2016)}]{Zhu.2016a}
Zhu T, Manhart M (2016) Oscillatory {{Darcy Flow}} in {{Porous Media}}.
  Transport in Porous Media 111(2):521--539. \doi{10.1007/s11242-015-0609-3}

\bibitem[{Zhu et~al(2014)Zhu, Waluga, Wohlmuth, and Manhart}]{Zhu.2014}
Zhu T, Waluga C, Wohlmuth B, et~al (2014) A {{Study}} of the {{Time Constant}}
  in {{Unsteady Porous Media Flow Using Direct Numerical Simulation}}.
  Transport in Porous Media 104(1):161--179. \doi{10.1007/s11242-014-0326-3}

\end{thebibliography}

\end{document}